\documentclass[]{raa}            
\usepackage{graphicx,times}
\usepackage{natbib}

\providecommand{\bm}[1]{\mbox{\boldmath$#1$\unboldmath}}

\begin{document}

   \title{Characteristics of convection and overshooting in RGB and AGB stars\footnote{Supported by the National Natural Science Foundation of China}
}

 \volnopage{ {\bf 2011} Vol.\ {\bf X} No. {\bf XX}, 000--000}
   \setcounter{page}{1}

   \author{X.-J. Lai
      \inst{1,2,3}
   \and Y. Li
      \inst{1,2}
   }

   \institute{National Astronomical Observatories/Yunnan Observatory, Chinese Academy of Sciences, Kunming 650011, China;
  {\it lxj@mail.ynao.ac.cn, ly@ynao.ac.cn}\\
        \and
             Key Laboratory for the Structure and Evolution of
             Celestial Objects, Chinese Academy of Sciences, Kunming 650011, China
        \and
             Graduate School of the Chinese Academy of Sciences, Beijing 100049,
             China.
\vs \no
   { }
}

\abstract{ Based on the turbulent convection model (TCM) of Li \&
Yang (2007), we have studied the characteristics of turbulent
convection in the envelopes of $2$ and $5M_{\odot}$ stars at the RGB
and AGB phases. The TCM has been applied successfully in the whole
convective envelopes including the convective unstable zone and the
overshooting regions. We find that the convective motions become
stronger and stronger when the stellar models are located upper and
upper along the Hayashi line. In the convective unstable zone, we
find that the turbulent correlations are proportional to functions
of a common factor $(\nabla-\nabla_{ad})\overline{T}$, which
explains similar distributions of those correlations. For the TCM we
find that if the obtained stellar structure of temperature is close
to that of the MLT, the convective motion will be at a much larger
velocity and thus more violent. However, if the turbulent velocity
is adjusted to close to that of the MLT, the superadiabatic
convection zone is much more extended inward and a lower effective
temperature of the stellar model will be obtained. For the
overshooting distance we find that the e-folding lengths of the
turbulent kinetic energy $k$ in both the top and bottom overshooting
regions decrease as the stellar model is located up along the
Hayashi line, but both the extents of the decrease are not obvious.
And the overshooting distances of the turbulent correlation
$\overline{u_{r}^{'}T^{'}}$ are almost the same for the different
stellar models with a same set of TCM's parameters. For the decay
way of the  kinetic energy $k$ we find that they are very similar
for the different stellar models based on a same set of TCM's
parameters, and there is a nearly linear relationship between $\lg
k$ and $\ln P$ for the different stellar models. When $C_{s}$ or
$\alpha$ increases while the other parameters are fixed, the
obtained linearly decaying distance will become longer.
 \keywords{RGB and AGB stars: convection of stars: turbulent
 convection model: convective overshooting
 } }

   \authorrunning{X.-J. Lai \& Y. Li }            
   \titlerunning{Characteristics of convection and overshooting in RGB and AGB stars }  
   \maketitle


%
%
\section{Introduction}           
\label{sect:intro} Red giant branch (RGB) and asymptotic giant
branch (AGB) stars are characterized by the convective motion in
their envelopes, which can extend over an enormous range in mass (or
radius). Because of huge scale of convection zones and small
viscosity of the stellar material, turbulent flow instead of laminar
one always occurs in their outer envelopes. Turbulent convection may
result in many important effects in stars: mixing different elements
to be homogeneous in the convection zone and adding fresh fuel to
the nuclear burning region, dredging the internal material processed
by H or He burning up to the stellar surface, and being as an
important means for heat transfer. These effects significantly
influence the structure and evolution of stars. Many observational
phenomena appearing in the RGB and AGB stars are ascribed to the
convection. For the RGB stars as an example, many of them show
chemical anomalies (Pilachowski et al. 1993; Charbonnel 1995;
Gratton et al. 2000; Kraft et al. 1993; Ram\'{\i}ez \& Cohen 2002;
Gratton et al. 2004; Busso et al. 2007; Recio-Blanco \& de Laverny
2007). And we are still perplexed to the so-called carbon star
mystery (Iben 1981; Iben 1975, 1977; Sackmann 1980; Lattanzio 1989;
Hollowell \& Iben 1988; Straniero et al. 1997; Herwig et al. 1997;
Herwig 2005) for the AGB stars. Furthermore, the observed mass loss
occurring in the RGB and AGB stars may be due to the turbulent
pressure (Jiang \& Huang 1997). To solve these problems we need to
accurately deal with the convection and to learn details of its
characteristics.
\parskip=.0in

For some RGB and AGB stars the convective motion near the surface of
the stellar envelope are superadiabtic because of less efficient
convective energy transport, and may sometimes become supersonic
based on the mixing-length theory (MLT) (Deng \& Xiong 2001). The
MLT (B\"{o}hm-Vitense 1953, 1958) is commonly used to deal with the
convection. However, due to its simplicity many shortcomings are
found (Renzini 1987; Baker \& Kuhfuss 1987; Spruit et al. 1990;
Pederson et al. 1990), which thus lead to some confused and
debatable problems. For example, convective overshooting (e.g.
Saslaw \& Schwarzschild 1965), semiconvection (e.g. Castellani et
al. 1971; Sreenivasan \& Wilson 1978) and the breathing convection
(e.g. Castellani et al. 1985).  Therefore new convection models have
been proposed successively, for example, non-local mixing length
theories (Ulrich 1970; Maeder 1975; Bressan et al. 1981) and
full-spectrum convection theory (Canuto \& Mazzitelli 1991).
However, they are still based on the framework of the MLT. A better
theory to overcome the MLT's drawbacks is the turbulent convection
models (TCMs), which are deduced directly from the Navier-Stokes
equations, and can describe many turbulence properties (Xiong 1979,
1981, 1985, 1990; Canuto 1992, 1994, 1997; Xiong, Cheng \& Deng
1997; Canuto \& Dubovikov 1998; Canuto et al. 1996). However, the
dynamic equations for turbulent correlations are expressed by higher
order correspondents due to the inherent non-linearity of the
Navier-Stokes equations. Therefore these equations are needed to be
cut-off with some closure approximations in order to be applicable
in calculations of the stellar structure and evolution.
Nevertheless, some free parameters have to be introduced by the
closure assumptions. Different closure assumptions correspond to
different TCMs, and many of them may not be easy to be incorporated
into a stellar evolution code. Recently a simple TCM proposed by Li
\& Yang (2007) was successfully applied to the solar models (Yang \&
Li 2007), in which the introduced parameters of the TCM are directly
related to the corresponding turbulence physics. Some significant
changes in the structure of the solar convection zone and better
results compared to the solar p-mode observations are obtained.
Shortly afterwards, Zhang \& Li (2009) applied it successfully to
the overshooting region of the solar convection zone.
\parskip=.0in

Up to now, the TCMs have seldom been applied to very large
convective envelopes of stars, e.g. in the RGB or AGB stars, which
are rather different from the solar environment. Furthermore, in
view of the key roles of convection played in these stars, we choose
three sites of stellar evolution to test the TCM of Li \& Yang
(2007). We try to find out convection characteristics for two stars
of 5 and $2M_{\odot}$ at the RGB and AGB phases, and to find out
their dependence on the TCM's parameter. To analyze the properties
of the turbulent correlations among velocity and temperature
fluctuations, we separate the whole convection envelope (derived
self-consistently from the TCM) into the convective unstable zone
and the convective overshooting regions, respectively. We firstly
describe the basic equations of the TCM in Section 2. The
information of the evolutionary code and input physics are described
in Section 3. In Section 4, the convection characteristics in the
convective unstable zone and overshooting regions are presented and
discussed, respectively. The dependence of the structure of the
overshooting regions on the TCM's parameters is given in Section 5.
Finally, some concluding remarks are summarized in Section 6.


\section{EQUATIONS OF TCM}
\label{sect:Obs}

Using the Reynolds decomposition approximation, and with the aid of
some closure approximations, the second-order moment equations of
the TCM are (Li \& Yang 2007 for details):

\begin{eqnarray}
&&\frac{1}{\overline{\rho} r^{2}}\frac{\partial}{\partial
\ln\overline{P}}\biggl(C_{s}\overline{\rho}r^{2}\frac{\overline{u_{r}^{'}u_{r}^{'}}}{\sqrt{k}}\frac{\partial\overline{u_{r}^{'}u_{r}^{'}}}{\partial
\ln\overline{P}}
\biggl)=C_{k}\frac{\sqrt{k}}{\alpha}\biggl(\overline{u_{r}^{'}u_{r}^{'}}-\frac{2}{3}k\biggl)+\frac{2}{3}\frac{k^{3/2}}{\alpha}+\frac{2\beta
g_{r}}{\overline{T}}H_{P}\overline{u_{r}^{'}T'},
 \label{eq:LebsequeI}
 \\
&&\frac{1}{\overline{\rho} r^{2}}\frac{\partial}{\partial
\ln\overline{P}}\biggl(C_{s}\overline{\rho}r^{2}\frac{\overline{u_{r}^{'}u_{r}^{'}}}{\sqrt{k}}\frac{\partial
k}{\partial \ln\overline{P}}
\biggl)=\frac{k^{3/2}}{\alpha}+\frac{\beta
g_{r}}{\overline{T}}H_{P}\overline{u_{r}^{'}T'},
 \label{eq:LebsequeII}
 \\
&&\frac{2}{\overline{\rho} r^{2}}\frac{\partial}{\partial
\ln\overline{P}}\biggl(C_{t1}\overline{\rho}r^{2}\frac{\overline{u_{r}^{'}u_{r}^{'}}}{\sqrt{k}}\frac{\partial
\overline{u_{r}^{'}T'}}{\partial \ln\overline{P}} \biggl)=
\biggl[\frac{\overline{\rho}c_{p}H_{P}}{\lambda}\overline{u_{r}^{'}T'}-\overline{T}(\nabla_{R}-\nabla
ad)\biggl]\overline{u_{r}^{'}u_{r}^{'}}+
\nonumber\\
&&\hspace{148pt}\frac{\beta
g_{r}}{\overline{T}}H_{P}\overline{{T'}^{2}}
+C_{t}\biggl(\frac{\sqrt{k}}{\alpha}+\frac{\lambda}{\overline{\rho}c_{p}\alpha^{2}
H_{P}}\frac{\varepsilon^{2}}{k^{3}}\biggl)\overline{u_{r}^{'}T'},
 \label{eq:LebsequeIII}\\
&&\frac{1}{\overline{\rho} r^{2}}\frac{\partial}{\partial
\ln\overline{P}}\biggl(C_{e1}\overline{\rho}r^{2}\frac{\overline{u_{r}^{'}u_{r}^{'}}}{\sqrt{k}}\frac{\partial
\overline{{T'}^{2}}}{\partial \ln\overline{P}} \biggl)=2
\biggl[\frac{\overline{\rho}c_{p}H_{P}}{\lambda}\overline{u_{r}^{'}T'}-\overline{T}(\nabla_{R}-\nabla
ad)\biggl]\overline{u_{r}^{'}T^{'}}+
\nonumber\\
&&\hspace{145.5pt}
2C_{e}\biggl(\frac{\sqrt{k}}{\alpha}+\frac{\lambda}{\overline{\rho}c_{p}\alpha^{2}
H_{P}}\biggl)\overline{{T'}^{2}}.
 \label{eq:LebsequeV}
\end{eqnarray}

The terms on the left-hand side of Eqs.(1)-(4) are diffusion ones of
the turbulence, representing the non-local characteristics of
turbulence and describing mainly the properties of the overshooting
regions outside the Schwarzschild boundaries of a convection zone.
The terms on the right-hand side of Eqs.(1)-(4) are some production
and dissipation ones of the turbulence.
 All the symbols appeared in above equations have their usual
meaning (Li \& Yang 2001; Li \& Yang 2007). There are seven free
parameters in above equations. They are $C_{t}$, $C_{e}$, $C_{k}$,
$C_{t1}$, $C_{e1}$, $C_{s}$, and $\alpha$. The former three ones are
dissipation parameters which describe respectively the dissipation
of the turbulent correlation $\overline{u_{r}^{'}T^{'}}$, the
dissipation of the temperature fluctuation $\overline{{T^{'}}^{2}}$,
and actually the anisotropic degree of the turbulence. The larger
the values of them are, the smaller the convective heat fluxes will
be (Li \& Yang 2007). The next three parameters are diffusion ones
related respectively to the three turbulent correlations
$\overline{u_{r}^{'}T^{'}}$, $\overline{{T^{'}}^{2}}$ and
$\overline{u_{r}^{'}u_{r}^{'}}$. Larger values of them will result
in lowered and expanded profiles of these correlations (Li \& Yang
2007). The last parameter $\alpha$, the ratio of a typical length
$l$ to the local pressure scaleheight, is introduced by the closure
assumption similar to the mixing length parameter in the MLT. The
effects of these parameters on the properties of turbulence, and
then on the structure of the solar convection zone, are investigated
by Li \& Yang (2007) and Zhang \& Li (2009). Here we choose several
sets of these parameters' values to study the properties of the
turbulent convection in the RGB and AGB stars.

\section{input physics}
We calculate the evolution of two stars of $2$ and $5M_{\odot}$
respectively. The initial composition (Z=0.02, X=0.7) is fixed for
both stellar models.

The free parameters adopted in the TCM can be determined by the
fluid dynamics experiments in the terrestrial environment (e.g.
Hossain \& Rodi 1982). Given a set of appropriate values of
dissipation and diffusion parameters, the value of $\alpha$ can be
determined by calibrating the solar model (Li \& Yang 2007; Zhang \&
Li 2009). The value of $\alpha$ obtained in this way can be used to
RGB or AGB calculations, but this approach is at the sacrifice of
the uncertainty of its value (Herwig 2005). In this paper, we adopt
several sets of the parameters' values, which are derived directly
from or mainly based on the works of Li \& Yang (2007) and Zhang \&
Li (2009) for the solar model and are shown in Table 1. In which the
values of $C_{t}$, $C_{e}$ and $C_{k}$ are mainly derived from the
terrestrial experiments and their typical values are 3.0, 1.25 and
2.5, respectively. Nevertheless, $C_{t}=7.0$ is taken from the work
of Canuto (1993) and $C_{t}=0.20, C_{e}=0.10$ from the suggestion
that smaller values of them will lead to a better result compared to
the solar p-mode observation (Yang \& Li 2007). Furthermore,
$C_{t1}=C_{e1}=10^{-7}$ are adopted, meaning a negligible diffusion
effect of $\overline{u_{r}'T'}$ and $\overline{T'^{2}}$ and a local
convection model of them, to assure that an almost same temperature
structure in stellar model can be obtained for both the MLT and TCM
and then a meaningful comparison between them can be done, which
will be discussed in the following subsection 4.3. For simplicity we
call the TCM with the diffusion terms in Eqs.(1)-(4) as NLMs and the
TCM without the diffusion terms as LM. The MLT with $\alpha=1.70$ is
also adopted to calculate the stellar convection at the RGB and AGB
phases for comparisons. In Table 1 both the parameter sets of NLMa
and NLMd are corresponding to the standard solar model calibration.

The stellar evolution code used in the present paper was originally
described by B. Paczynski and M. Kozlowski and updated by R.
Sienliewicz. Nuclear reaction rates are adopted from BP95 (Bahcall,
Pinsonneault \& Wasserburg 1995). The equation of state is the OPAL
EOS from Rogers (1994) and Rogers, Swenson \& Iglesias (1996). The
OPAL opacities GN93hz series (Rogers \& Iglesias 1995; Iglesias \&
Rogers 1996) are used in the high-temperature region. In the stellar
atmosphere, low-temperature opacities from Alexander \& Ferguson
(1994) are used.

To study the convection properties based on the TCM, we select three
evolved models during the RGB and AGB phases derived from the MLT,
which are indicated by solid stars in Fig.1.

\begin{table}
\bc
\begin{minipage}[]{100mm}
\caption[]{ The information of the TCM's parameters}\end{minipage}
\small
 \begin{tabular}{ccccccccccc}
  \hline\noalign{\smallskip}
Model Name & $C_{t}$  &  $C_{e}$   &    $C_{k}$   &    $C_{t1}$   &    $C_{e1}$   &    $C_{s}$   &    $\alpha$\\
  \hline\noalign{\smallskip}
MLT    &    &    &    &    &    &    &    $1.70$ \\
LM   & $3.0$  & $1.25$  & $2.5$   &    &    &   & $0.90$ \\
NLMa   & $3.0$  & $1.25$  & $2.5$   & $0.03$    &$0.03$    &$0.03$   & $0.90$ \\
NLMb1   & $7.0$  & $0.20$  & $2.5$   & $1.0\times10^{-7}$    & $1.0\times10^{-7}$    &$0.10$   & $0.76$ \\
NLMb2   & $7.0$  & $0.20$  & $2.5$   & $1.0\times10^{-7}$    & $1.0\times10^{-7}$    &$0.10$   & $0.74$ \\
NLMb3   & $7.0$  & $0.20$  & $2.5$   & $1.0\times10^{-7}$    & $1.0\times10^{-7}$    &$0.10$   & $0.15$ \\
NLMc1   & $3.0$  & $1.25$  & $2.5$   & $0.05$    &$0.05$    &$0.05$   & $0.68$ \\
NLMc2   & $3.0$  & $1.25$  & $2.5$   & $0.05$    &$0.05$    &$0.04$   & $0.68$ \\
NLMc3   & $3.0$  & $1.25$  & $2.5$   & $0.05$    &$0.05$    &$0.03$   & $0.68$ \\
NLMc4   & $3.0$  & $1.25$  & $2.5$   & $0.05$    &$0.05$    &$0.02$   & $0.68$ \\
NLMc5   & $3.0$  & $1.25$  & $2.5$   & $0.05$    &$0.05$    &$0.01$   & $0.68$ \\
NLMd   & $0.2$  & $0.10$  & $2.5$   & $0.15$    &$0.25$    &$0.10$   & $0.05$ \\
  \noalign{\smallskip}\hline
\end{tabular}
\ec
\end{table}

   \begin{figure}[h!!!]
   \centering
   \includegraphics[width=7.0cm, angle=0]{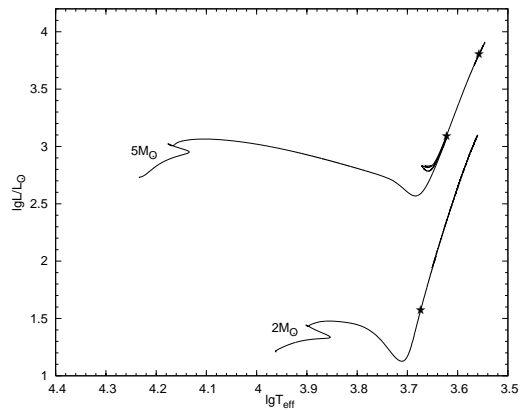}
   \begin{minipage}[]{135mm}
   \caption{The HR diagram for the stellar models of $2$ and $5M_{\odot}$ based on the MLT. The three solid stars indicate the certain locations of the stellar models during the RGB $(2M_{\odot}: \lg L/L_{\odot}=1.58, T_{eff}=4750K; 5M_{\odot}: \lg L/L_{\odot}=3.09, T_{eff}=4190K)$ and AGB ($5M_{\odot}: \lg L/L_{\odot}=3.81, T_{eff}=3570K)$ phases} \end{minipage}
   \label{Fig1}
   \end{figure}

   \begin{figure}[h!!!]

    \begin{minipage}[t]{70mm}
    \centering

   \includegraphics[width=6.0cm,height=8.5cm,angle=0]{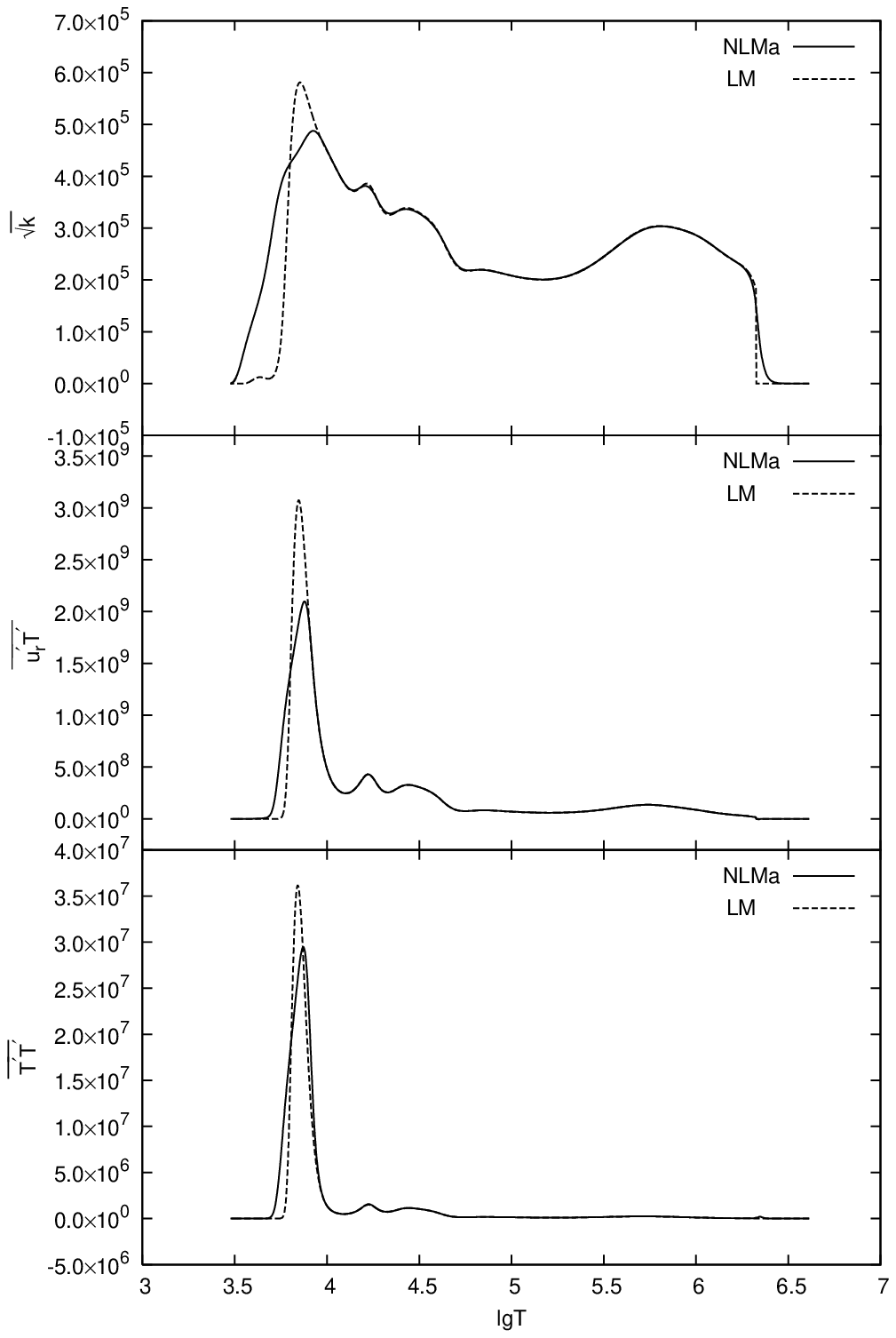}

   \caption{Distributions of the turbulent correlations for the $5M_{\odot}$ star at the AGB phase. The solid lines are for NLMa, dashed lines for LM. }
   \label{Fig2}
   \par\vspace{10pt}
   \end{minipage}%
    \begin{minipage}[t]{70mm}
    \centering
   \includegraphics[width=6.0cm,height=8.5cm, angle=0]{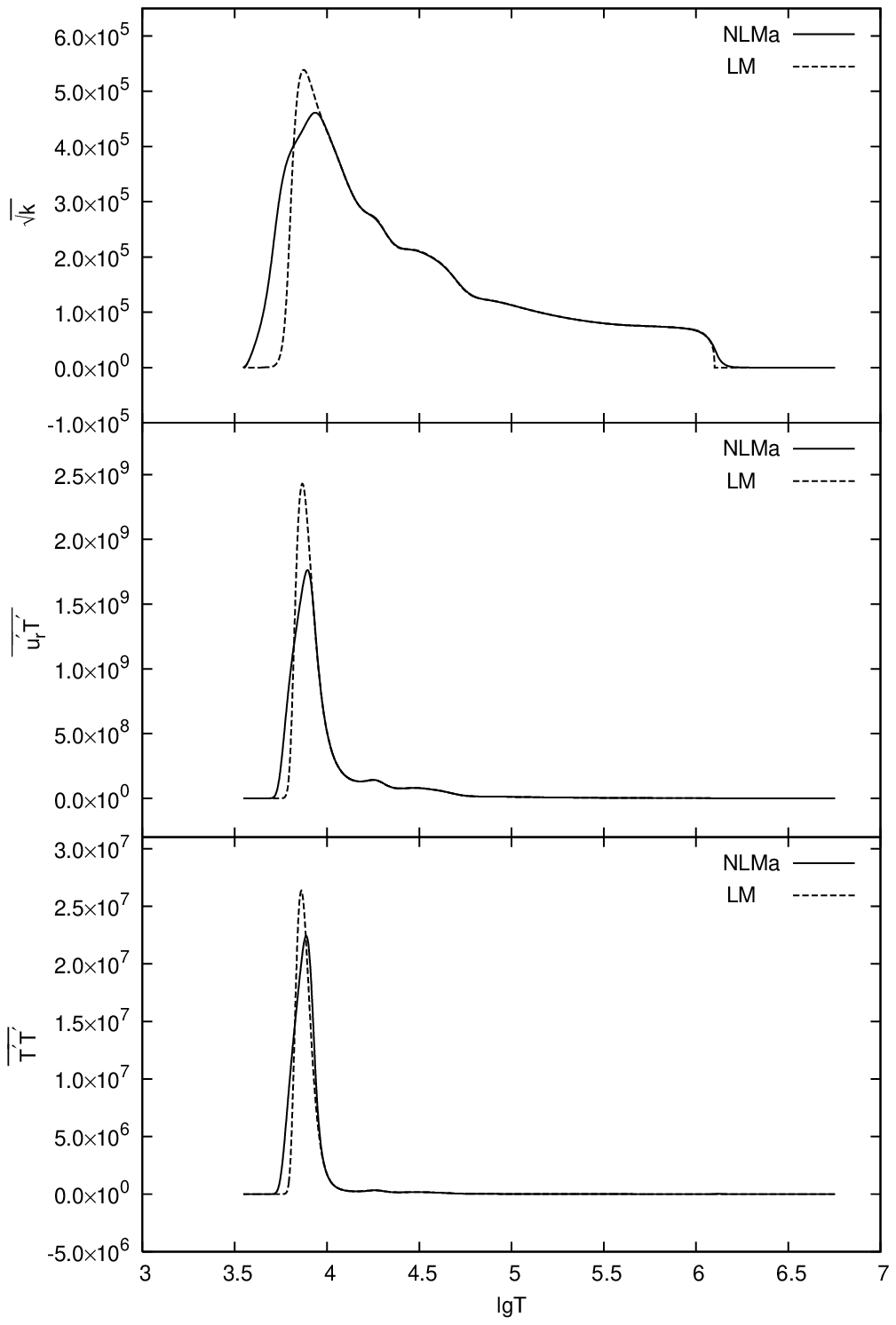}
   \caption{The same as Fig.2, but for the $5M_{\odot}$ star at the RGB phase. }
   \label{Fig3}
\end{minipage}
   \end{figure}

   \begin{figure}[h!!!]
   \centering
   \vspace{-25pt}
   \includegraphics[width=6.0cm,height=8.5cm, angle=0]{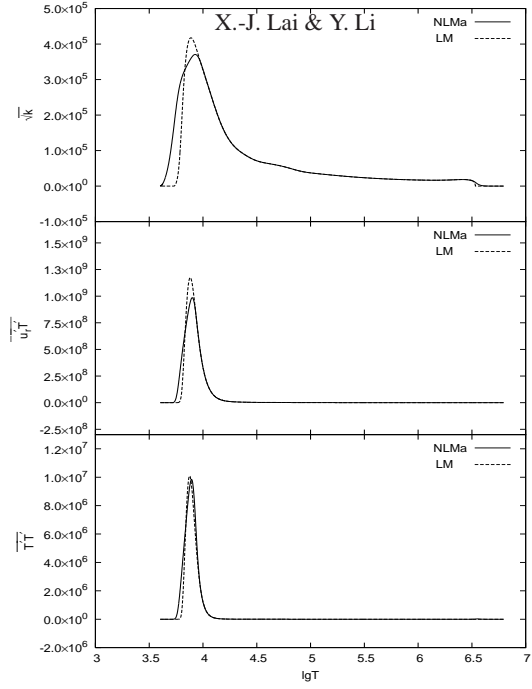}
   \begin{minipage}[]{95mm}
   \
   \caption{The same as Fig.2, but for the $2M_{\odot}$ star at the RGB
   phase.}
   \end{minipage}
   \label{Fig4}
   \end{figure}
\section{Characteristics of turbulence}
\subsection{Profiles of the Turbulent Correlations}

The profiles of the turbulent correlations $\sqrt{k}$,
$\overline{u_{r}^{'}T^{'}}$ and $\overline{{T'}^{2}}$ are displayed
in Fig.2 for the $5M_{\odot}$ model at the AGB phase. It can be seen
that convection widely develops in the stellar envelope $(3.5\leq
\lg T\leq6.4)$ and the turbulent kinetic energy $k$ maintains at a
high level in most of the convection zone. Particularly, there are
several peaks for all the three correlations appearing at almost the
same locations. The appearance of the maximum turbulent kinetic
energy is mainly due to ionization of hydrogen and helium which
intensively blocks the transfer of heat and results in strong
buoyancy to drive the convective motion. The profiles of the
correlations $\overline{u_{r}^{'}T^{'}}$ and $\overline{{T'}^{2}}$
vary in a similar way as that of $\sqrt{k}$, except that the largest
peaks are much higher than that of $\sqrt{k}$.

Diffusion terms in the turbulence equations are taken into account
in NLMa but not in LM. As a result the non-local effect is behaved
remarkably for NLMa around the boundaries of the convective unstable
zone, especially around the upper boundary where the NLMa's solution
significantly deviates from the LM's. However, the non-local effect
can be safely ignored in the interior of the convective unstable
zone because the turbulent correlations distribute quite
homogeneously there. These conclusions agree well with those of Yang
\& Li (2007) and Deng \& Xiong (2008).

However, for the other two stellar models of $5$ and $2M_{\odot}$ at
the RGB phase, the profiles of the three correlations, which are
respectively shown in Fig.3 and Fig.4, have some differences for
both NLMa and LM. These two models have smaller values of the three
correlations. It means that the turbulent motion is less violent and
thus transfers less heat. Furthermore, except for the maximum peak
of those correlations the other peaks become less prominent,
especially for the model of $2M_{\odot}$ at the RGB phase. Besides,
near the bottom of the convection envelope the turbulent kinetic
energy $k$ has a weak peak, and its value seems to decrease
monotonically with the stellar luminosity. In another word, the
convective motion becomes stronger and stronger when the stellar
models are located up along the Hayashi line.
\subsection{Turbulence Properties in the Convective Unstable Zone}
In order to understand the properties of turbulence in the
convective unstable zone, some assumptions are invoked to obtain the
explicit relationships between the correlations $\sqrt{k}$,
$\overline{u_{r}^{'}T^{'}}$ and $\overline{{T'}^{2}}$. According to
the result stated in the former section, the diffusion terms on the
left hand sides of Eqs.(1)--(4) can be approximately ignored in the
convective unstable zone. With the aid of Eqs.(1)--(4) we get:
\begin{equation}
\overline{{T'}^{2}}=\frac{D(\nabla-\nabla_{ad}){\overline{T}}^{2}k-D(\nabla-\nabla_{ad}){\overline{T}}^{2}k\sqrt{1+4\frac{C_{t}}{C_{e}}\frac{k}{D^{2}\alpha^{2}H_{P}g_{r}\beta(\nabla_{ad}-\nabla)}}}{2H_{P}g_{r}\beta},
\end{equation}
where $D=\frac{4}{3C_{k}}+\frac{2}{3}$. In view of the mixing-length
theory:
\begin{equation}
k=\frac{g\delta}{8H_{P}}(\nabla-\nabla_{ad})l^{2}\sim
c_{p}\nabla_{ad}(\nabla-\nabla_{ad})\overline{T},
\end{equation}
Eq.(5) can be further reduced to be:

\begin{equation}
\overline{{T^{'}}^{2}}\sim\frac{D\alpha^{2}}{8}(\nabla-\nabla_{ad})^{2}{\overline{T}}^{2}.
\end{equation}
By utilizing Eq.(6) and Eq.(2), the velocity-temperature correlation
can be approximated as:
\begin{equation}
\overline{u_{r}^{'}T^{'}}\sim\sqrt{c_{p}\nabla_{ad}}(\nabla-\nabla_{ad})^{3/2}{\overline{T}}^{3/2}.
\end{equation}

It can be seen that all the three correlations depend on
$(\nabla-\nabla_{ad}){\overline{T}}$, which explains why the three
correlations peak at almost the same places in the convective
unstable zone. Fig.5 shows the profiles of $k$,
${(\overline{u_{r}^{'}T^{'}})}^{2/3}$,
$({\overline{{T'}^{2}})}^{1/2}$ and
$(\nabla-\nabla_{ad}){\overline{T}}$ for NLMa of the $5M_{\odot}$
star at the AGB phase. It can be seen that the profiles of
${(\overline{u_{r}^{'}T^{'}})}^{2/3}$,
$({\overline{{T'}^{2}})}^{1/2}$ and
$(\nabla-\nabla_{ad}){\overline{T}}$ behave in a very similar way in
most of the convective unstable zone. However, the profile of $k$
deviates significantly from the others, but its overall trend is
almost consistent with them. It should be noted that the above
results are suitable for other stellar models.
   \begin{figure}[h!!!]
   \flushleft
   \qquad    \qquad    \quad
   \includegraphics[width=9.0cm, angle=0]{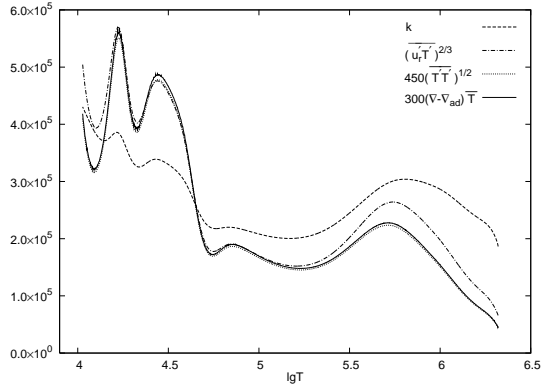}
   \begin{minipage}[]{135mm}
   \caption{Profiles of $k$,
${(\overline{u_{r}^{'}T^{'}})}^{2/3}$,
$({\overline{{T'}^{2}})}^{1/2}$ and
$(\nabla-\nabla_{ad}){\overline{T}}$ at the AGB phase  of the
$5M_{\odot}$ star based on NLMa.} \end{minipage}
   \label{Fig5}
   \end{figure}

\subsection{Comparisons with the MLT}
   \begin{figure}[h!!!]

   \centering

   \includegraphics[width=5.5cm,height=6.6cm]{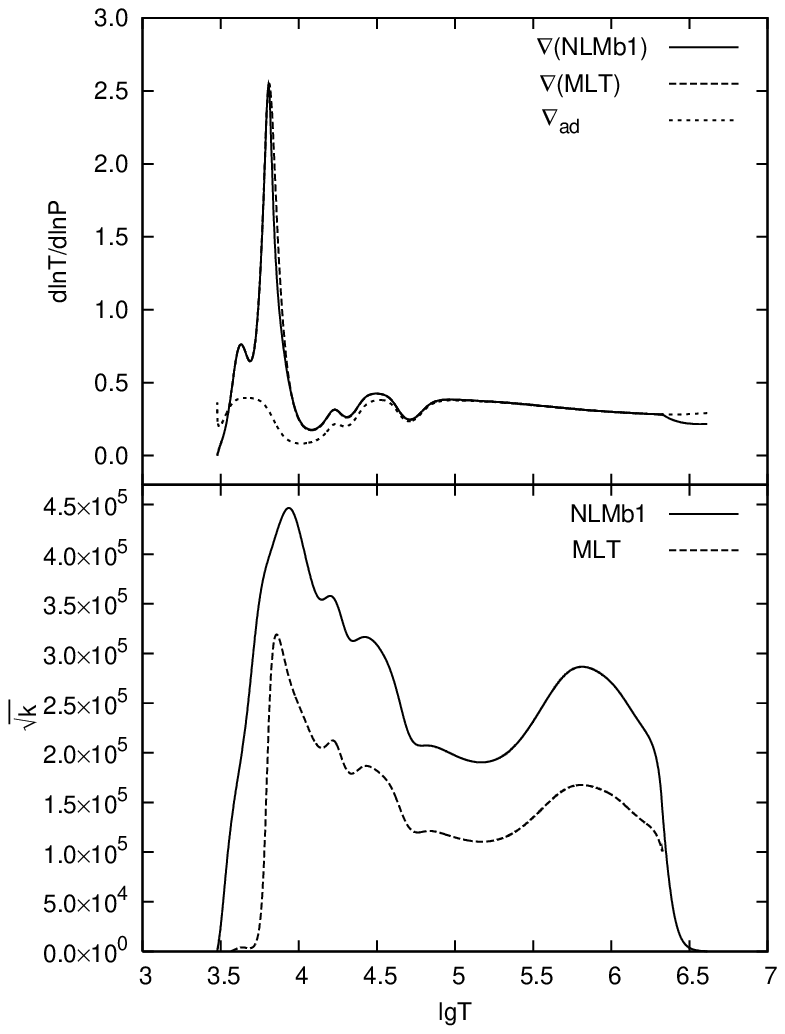}
   \qquad   \qquad   \quad
   \includegraphics[width=5.5cm,height=6.6cm]{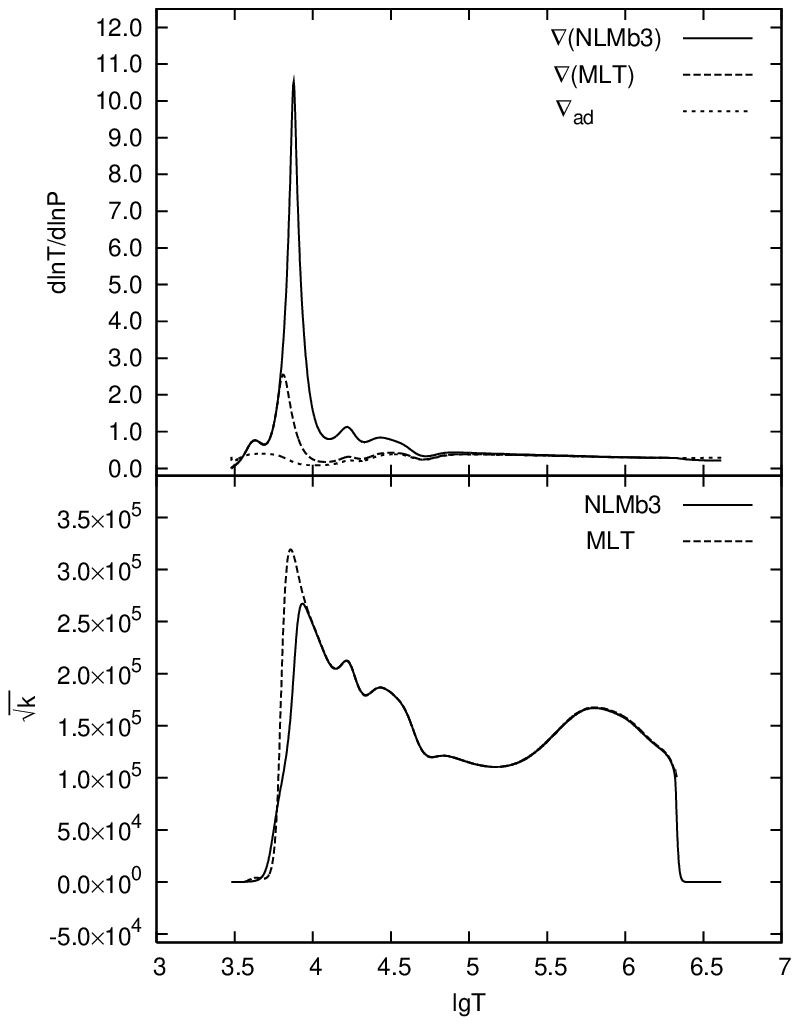}

   \

   \caption{The temperature gradient (including the real, radiative and adiabatic ones) and the square root of the turbulent energy $\sqrt{k}$ as functions of the temperature for the $5M_{\odot}$ star at the AGB phase (the same as in Fig.2). The dashed line is for MLT, the solid lines in left panels are for NLMb1 and the solid lines in right panels are for NLMb3.}
   \label{Fig7}
   \end{figure}

Results of two TCMs (NLMb1 and NLMb3) are compared with that of the
MLT. Firstly, for the comparison between NLMb1 and the MLT in the
left panel of Fig.6, we can find that the temperature gradient
$\nabla$ are almost the same with each other, which means that NLMb1
will result in a similar stellar structure as the MLT does in the
convective unstable zone. However, the value of the turbulence
velocity $(\sim\sqrt{k})$ of NLMb1 is much larger than  that of the
MLT. In another word, for a fixed heat flux transferred by the
convective motion, a larger value of convective velocity is
requested for the TCM. This is partly because that the dissipation
effects of the turbulent convection, measured by the parameters
$C_{t}$, $C_{e}$ and $C_{k}$, are fully taken into account.

When decreasing the value of parameter $\alpha$, for example
$\alpha=0.15$, however, we obtain an almost equivalent profile of
the convective velocity as the result of the MLT, which is shown in
the right panel of Fig.6 (namely NLMb3). It can be found that the
temperature gradient $\nabla$ of NLMb3 is much larger than that of
the MLT near the top of the convection envelope and extending to
$\lg T\approx4.7$ where it approaches the adiabatic temperature
gradient $\nabla_{ad}$, a place much deeper than the case of the MLT
being approximately at $\lg T\approx4.1$. It means that the
superadiabatic convection region is much more extended inward for
the TCM than the case of the MLT and the stellar models based on the
TCM will be of lower effective temperature than that of the MLT.

The results obtained above are mainly based on the star of
$5M_{\odot}$ at the AGB phase. Likewise, these conclusions are still
valid for the $2M_{\odot}$ star at the RGB phase and the
$5M_{\odot}$ star at the RGB phase. Fig.7, as an example, shows the
case of the $2M_{\odot}$ star at the RGB phase. It should be noted
from the right panel of Fig.7 that the superadiabatic convection
region does not extend so much inward as in the case of the
$5M_{\odot}$ star at the AGB phase.

   \begin{figure}[h!!!]

   \centering

   \includegraphics[width=5.8cm,height=6.6cm]{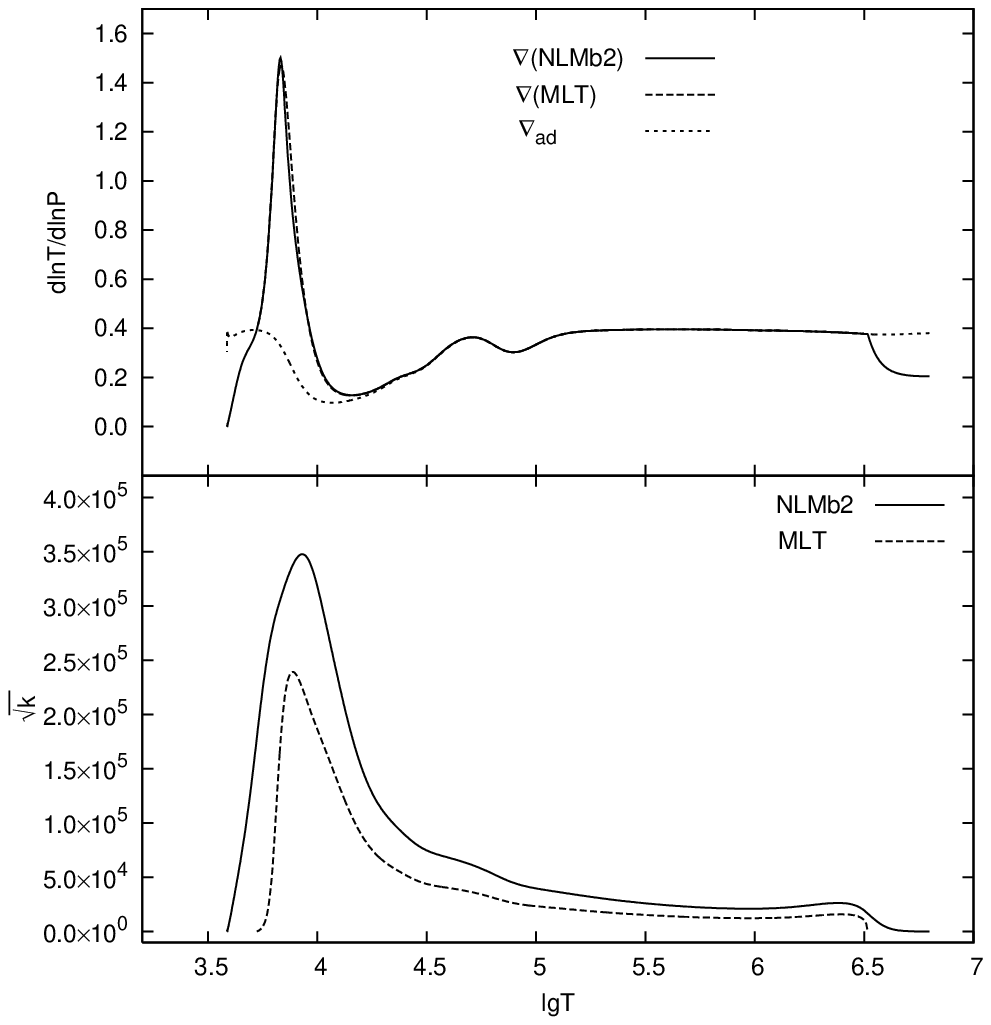}
   \qquad   \qquad
   \includegraphics[width=5.8cm,height=6.6cm]{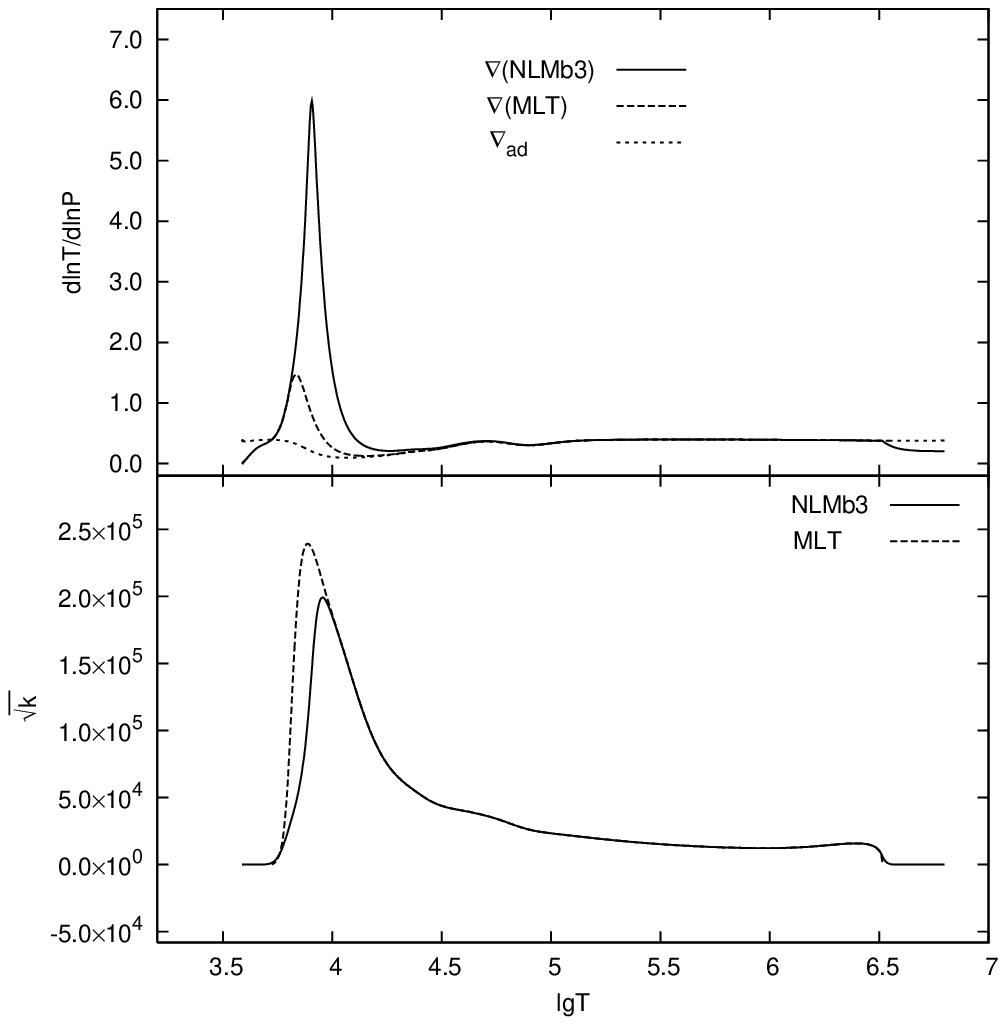}

   \

   \caption{The same as Fig.6, but for the $2M_{\odot}$ star at the RGB phase.}
   \label{Fig9}
   \end{figure}

\subsection{Turbulence Properties in the Overshooting Region}

 The turbulent correlations in the top and bottom overshooting regions are shown in Figs.8-10 for the considered stellar models. It can be found that the turbulent kinetic
 energy $k$ monotonously decreases outwards and tends to be zero from the boundaries of the convective unstable zone into the overshooting regions. For the
 correlation $\overline{{T'}^{2}}$, there is a peak in both the top and bottom overshooting regions, its value being positive and much larger than that of Zhang \& Li (2009). There is an exception in the top overshooting region for
 the $2M_{\odot}$ star at the RGB phase seen in the left panel of Fig.10, the peak
 of $\overline{{T'}^{2}}$ disappearing completely. Nevertheless, the value of the
 correlation $\overline{u_{r}^{'}T^{'}}$ is always negative in both the top and bottom overshooting regions. As a result, the temperature
 gradient $\nabla$ is greater than the radiative temperature
 gradient $\nabla_{R}$ but smaller than the adiabatic one
 $\nabla_{ad}$.

   \begin{figure}[h!!!]

    \begin{minipage}[t]{70mm}
    \centering
    \vspace{-30pt}
   \includegraphics[width=5.75cm,height=4.5cm,angle=0]{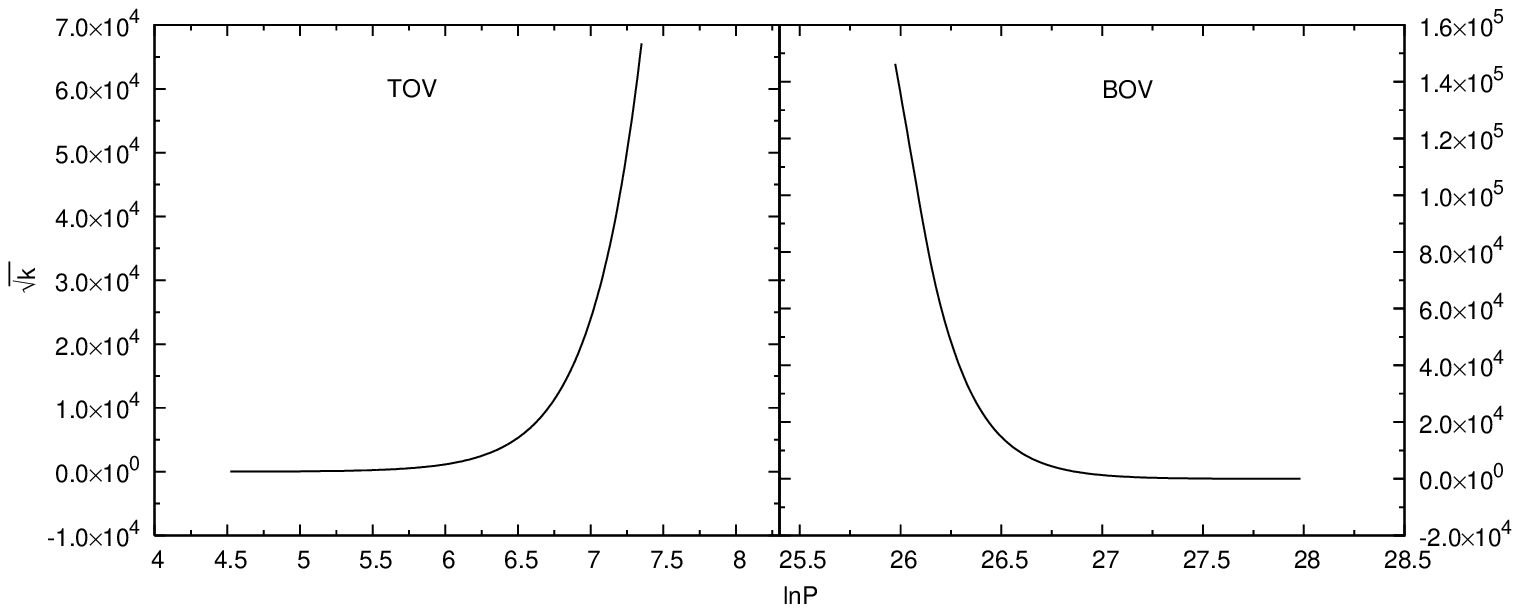}
   \par\vspace{-38pt}
   \includegraphics[width=5.75cm,height=4.5cm,angle=0]{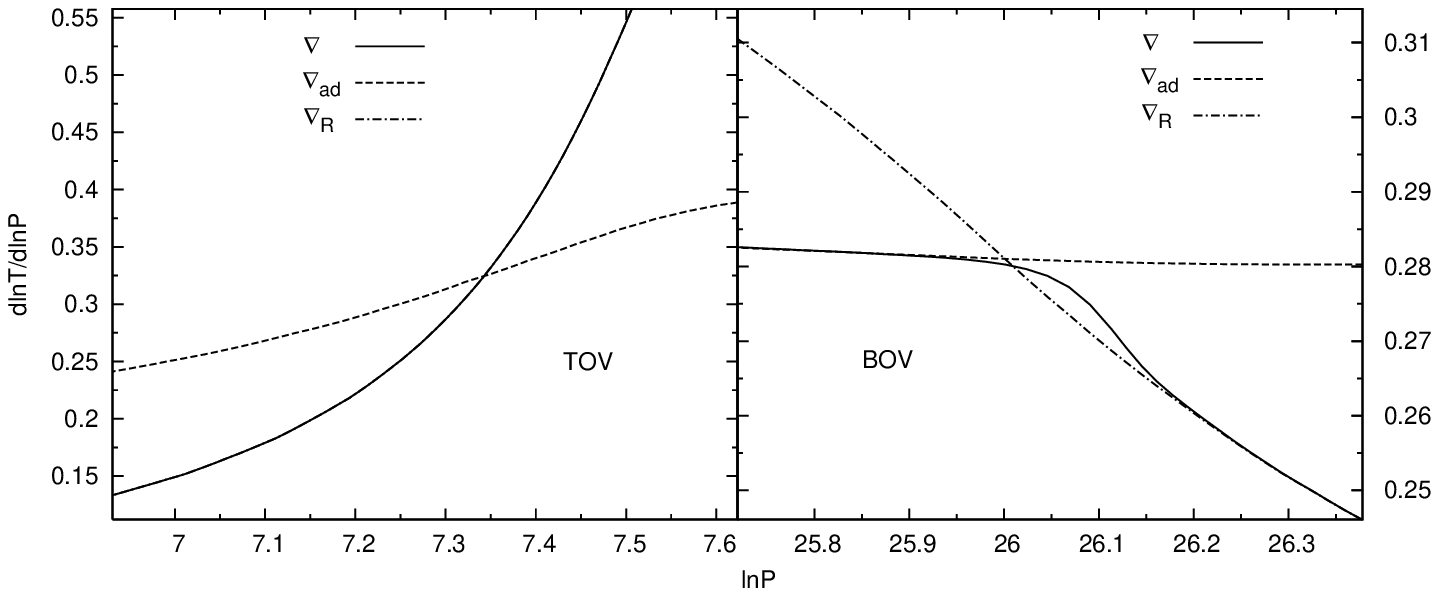}
   \par\vspace{-38pt}
   \includegraphics[width=5.75cm,height=4.5cm,angle=0]{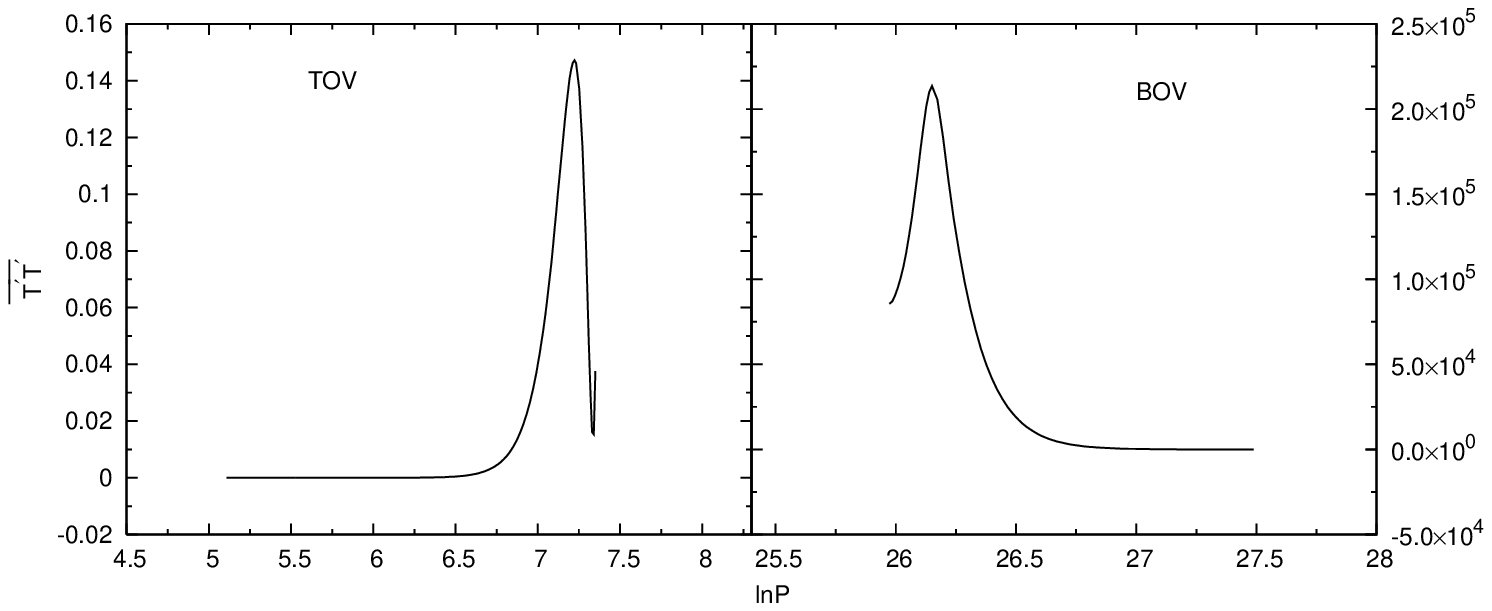}
   \caption{Distributions of $\sqrt{k}$, $\overline{{T'}^{2}}$ and the temperature gradients in the top overshooting region (TOV) and the bottom overshooting region (BOV) of the $5M_{\odot}$ star at the AGB phase according to NLMa.}
   \label{Fig12}
   \end{minipage}%
    \begin{minipage}[t]{70mm}
    \centering
    \vspace{-30pt}
   \includegraphics[width=5.75cm,height=4.5cm, angle=0]{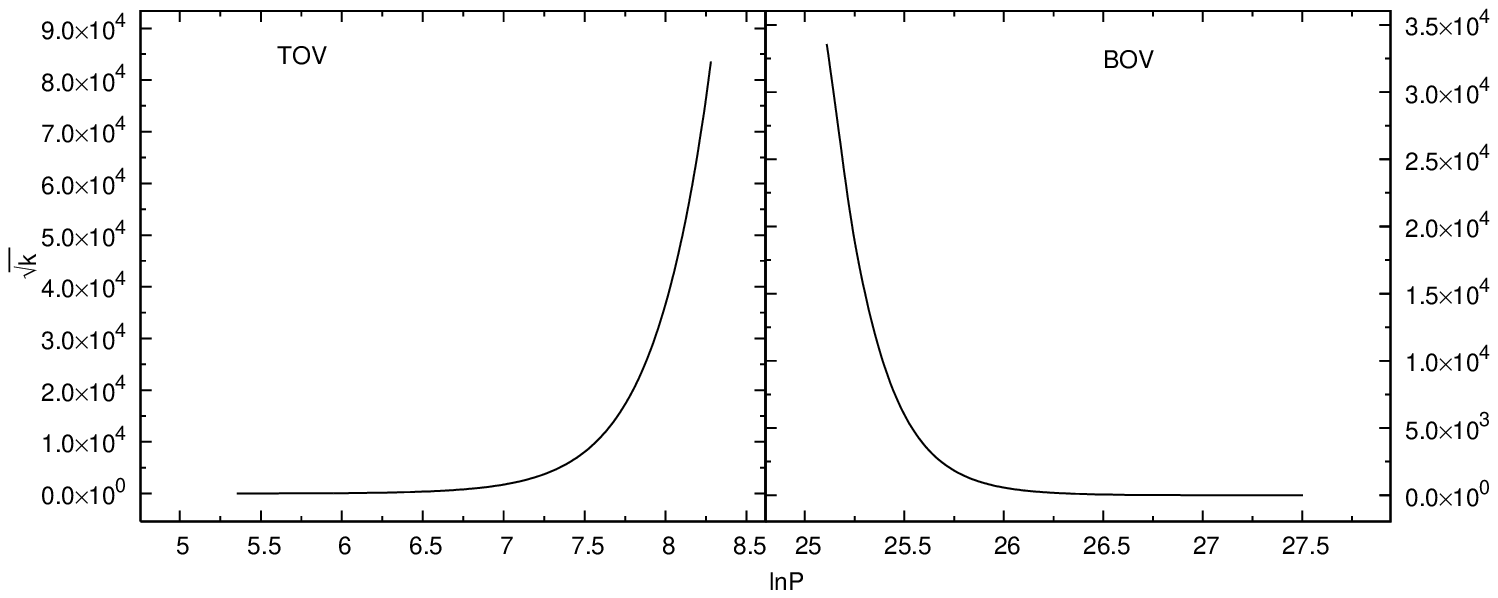}
   \par\vspace{-38pt}
   \includegraphics[width=5.75cm,height=4.5cm, angle=0]{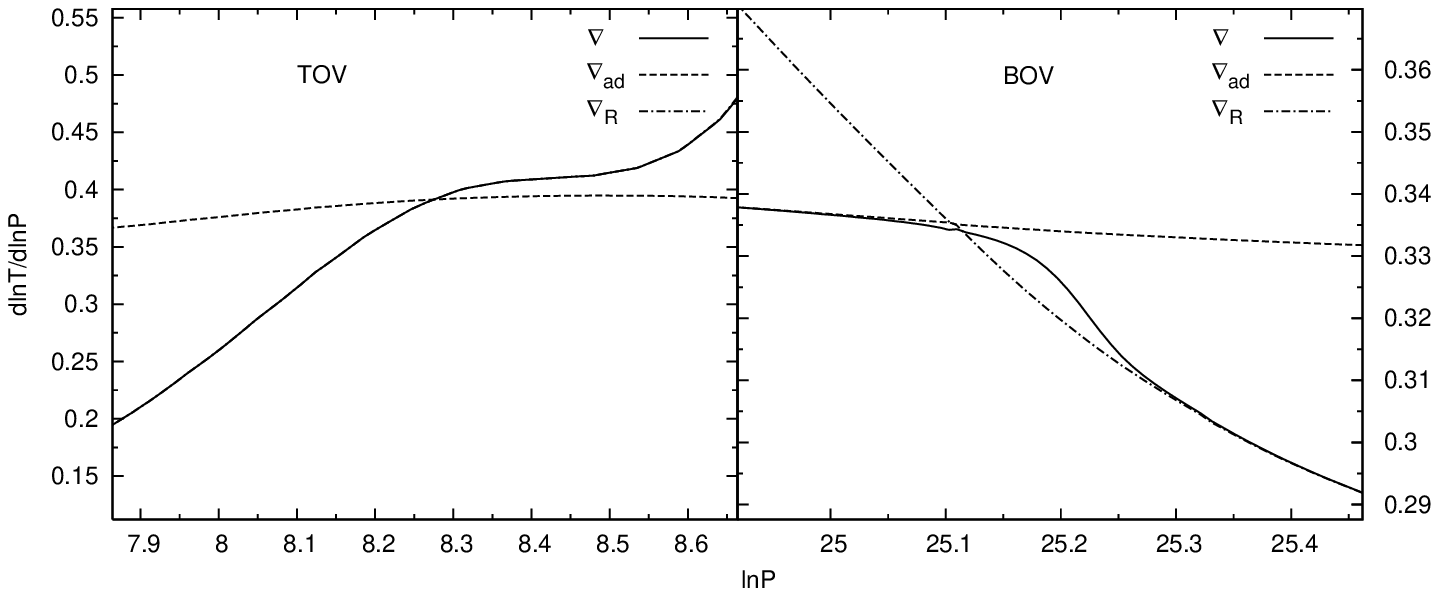}
   \par\vspace{-38pt}
   \includegraphics[width=5.75cm,height=4.5cm, angle=0]{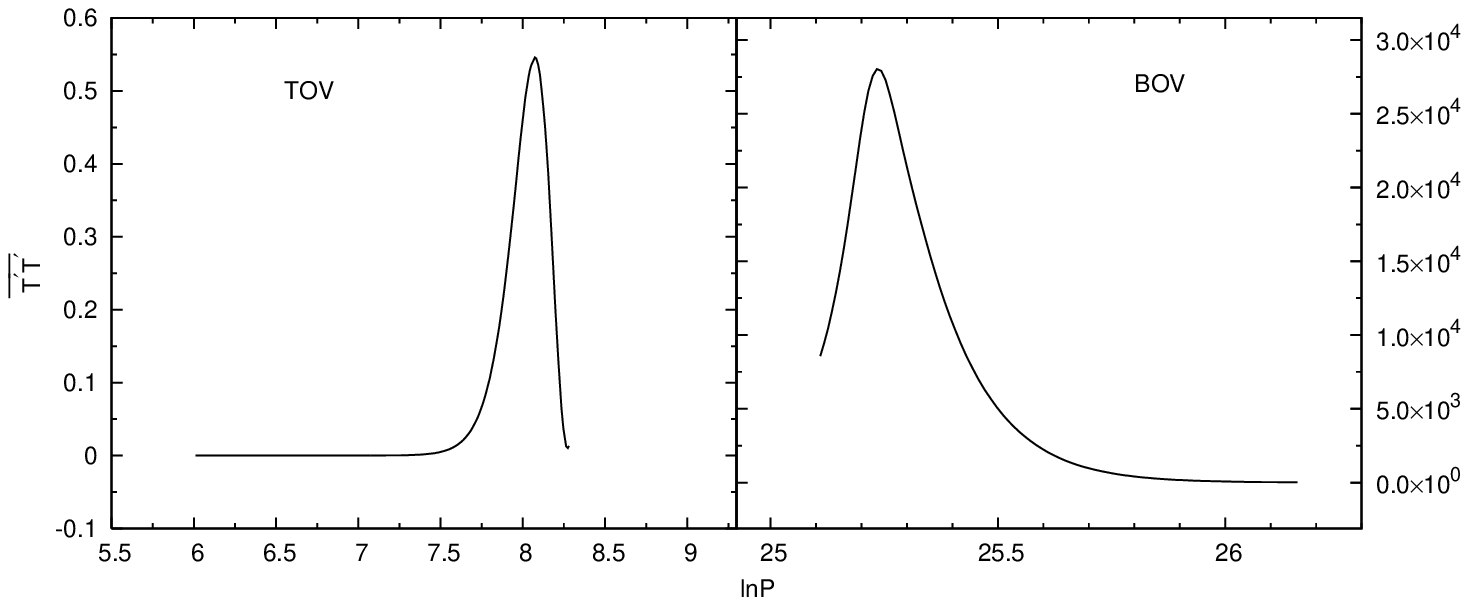}
   \caption{The same as Fig.8, but for the $5M_{\odot}$ star at the RGB phase. }
   \label{Fig9}
\end{minipage}
   \end{figure}
On the other hand, the overshooting extents are different for the
different correlations in a certain overshooting region. The profile
of $\sqrt{k}$ approximately demonstrates the distance that a
convective cell can reach beyond the boundaries of the convective
unstable zone. Its overshooting distance and decaying way will be
very important to the chemical mixing in the overshooting region, if
it is used to construct the diffusion process of chemical mixing
(e.g. Freytag et al. 1996, Herwig et al. 1997, Salasnich et al.
1999, Ventura \& D'Antona 2005). For the three stellar models, our
numerical results show that the e-folding length of $\sqrt{k}$ in
the top overshooting region decreases as the stellar model is
located upper and upper along the Hayashi line (they are 0.37, 0.34
and 0.34, respectively, in the unit of the local pressure
scaleheight $H_{P}$), and the case in the bottom overshooting region
behaves in the same way (they are 0.24, 0.23, 0.21$H_{P}$,
respectively), but the difference among them is not obvious in both
the top and bottom overshooting regions. However, for the case of
$\overline{u_{r}^{'}T'}$ that can be inferred by the length of the
region satisfying $\nabla_{ad}>\nabla>\nabla_{R}$, all the
overshooting distances in the bottom overshooting regions for the
three models are about $0.15H_{P}$. It means that the bottom
convective overshooting has a similar effect on the stellar
structure for all of the stellar models. In the top overshooting
regions, however, it can be seen that $\nabla$ and $\nabla_{R}$
nearly completely overlap with each other, except for a tiny
difference for the $2M_{\odot}$ star at the RGB phase. This may be
due to very low density in the top overshooting regions, resulting
in the convective heat flux comparatively rather low.

   \begin{figure}[h!!!]
   \centering
   \vspace{-30pt}
   \includegraphics[width=7.0cm, angle=0]{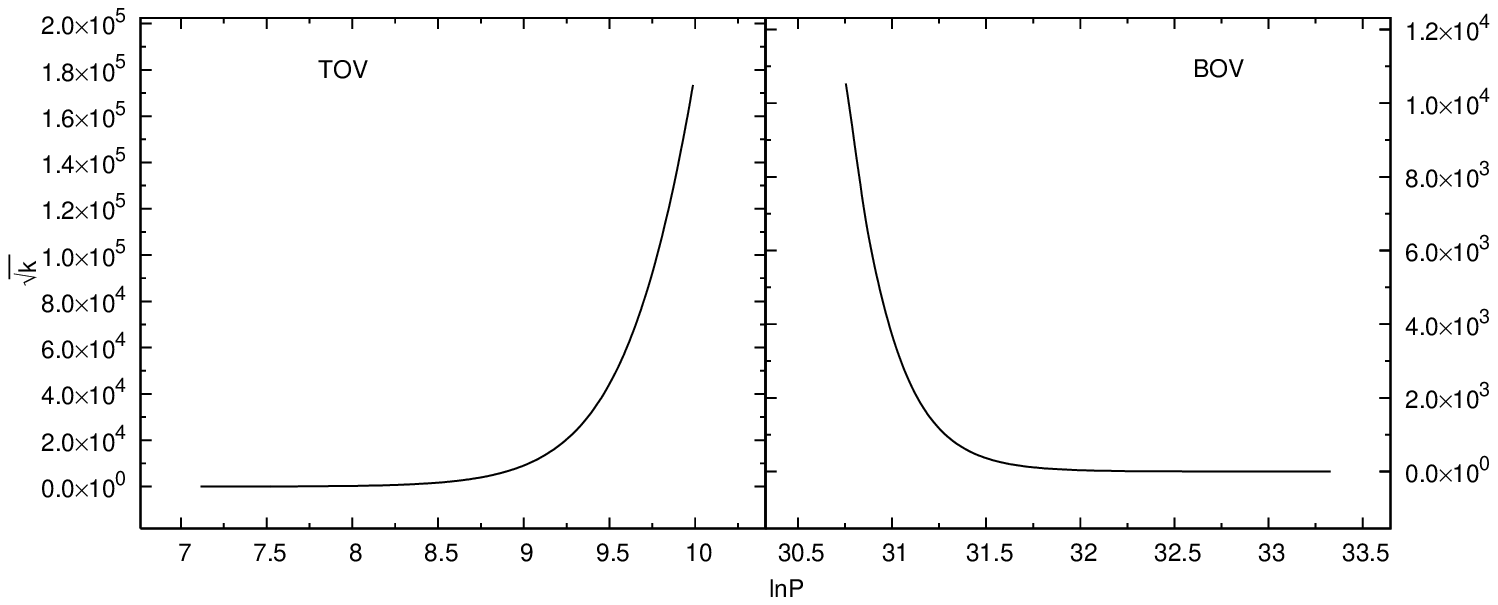}
   \par\vspace{-42pt}
   \includegraphics[width=7.0cm, angle=0]{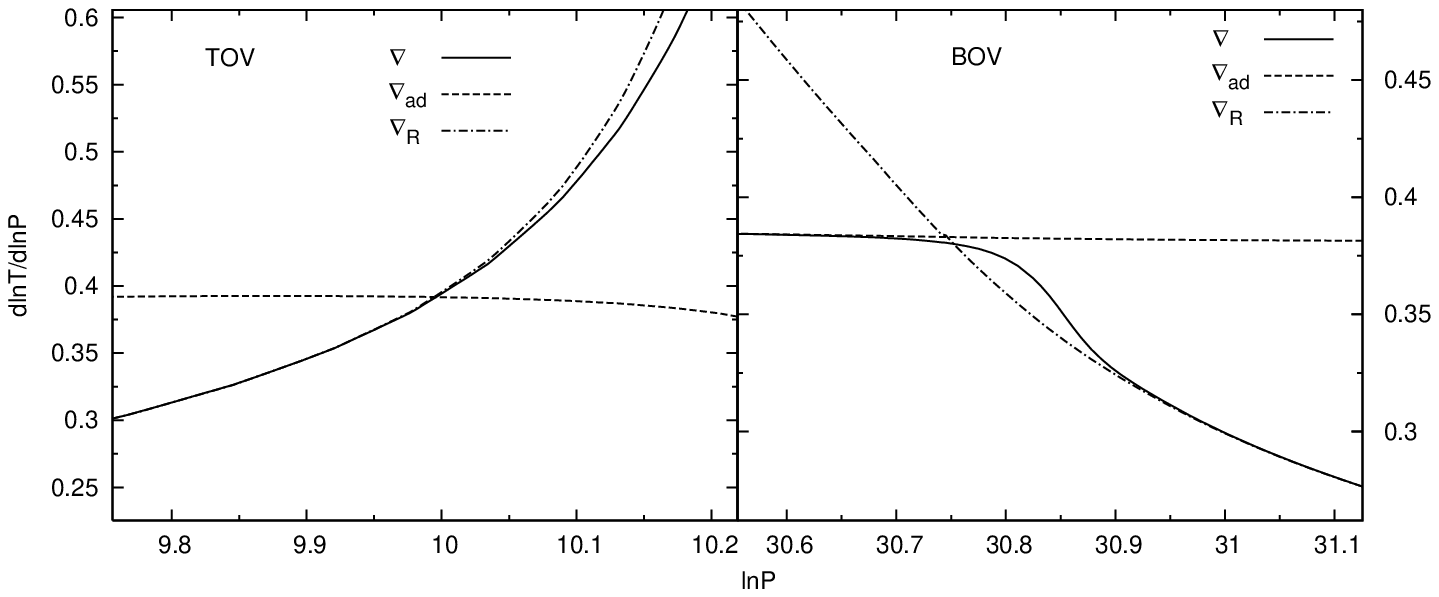}
   \par\vspace{-42pt}
   \includegraphics[width=7.0cm, angle=0]{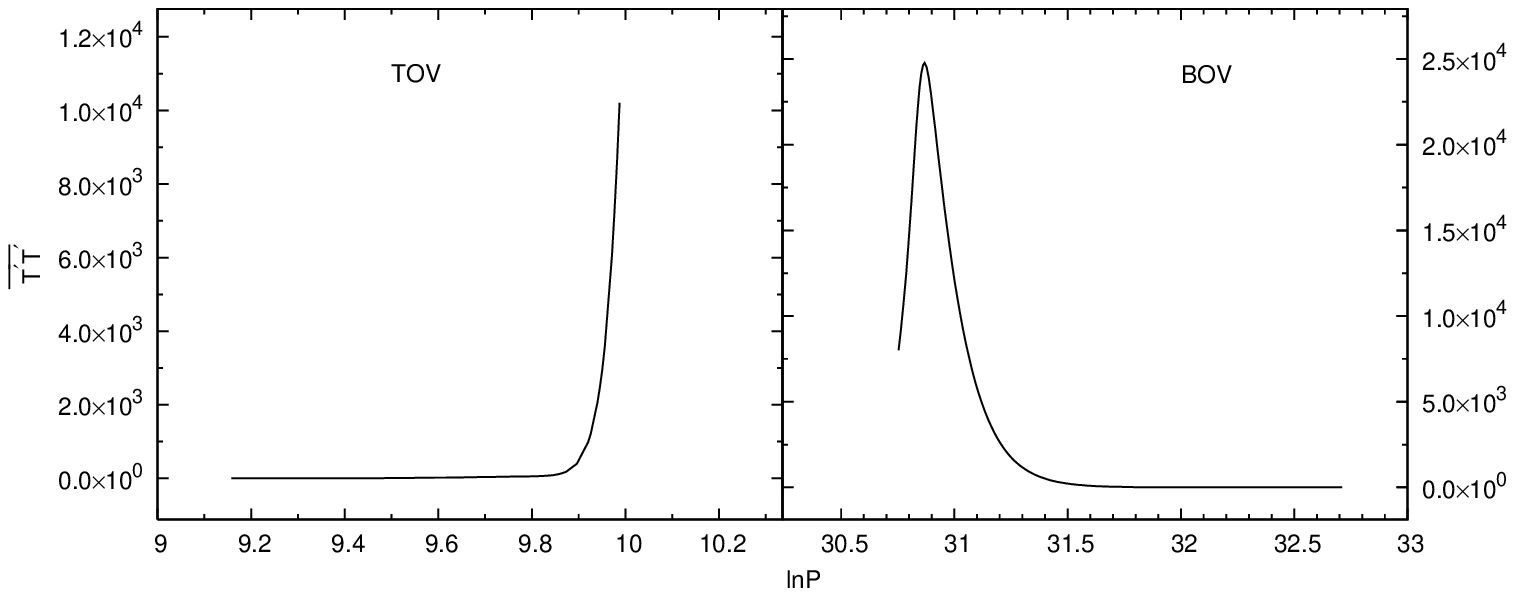}
   \begin{minipage}[]{135mm}
   \caption{The same as Fig.8, but for the $2M_{\odot}$ star at the RGB phase. }
   \end{minipage}
   \label{Fig18}
   \end{figure}

\section{Dependence of the structure of the overshooting regions on TCM's parameters}
 The dependence of convection characteristics of the sun on the TCM's parameters has been investigated extensively (Li \& Yang 2007; Zhang \& Li 2009). The turbulent correlations are mainly affected by the corresponding dissipation and diffusion parameters. The ratio of the radial kinetic energy to the horizontal
 one $\overline{u_{r}^{'}u_{r}^{'}}/\overline{u_{h}^{'}u_{h}^{'}}$ are shown in Fig.11 for the AGB model of
 $5M_{\odot}$, focusing on the effect of value of $C_{s}$ on the convective structure in the overshooting regions,
 because $C_{s}$ is the most sensitive parameter to determine the overshooting distance (Zhang \& Li 2009). It can be seen that with the convective cells penetrating more and more into the top overshooting region the convection becomes more and more horizontally dominated. Particularly, the turbulence is isotropic
 when $\overline{u_{r}^{'}u_{r}^{'}}/\overline{u_{h}^{'}u_{h}^{'}}=0.5$. It can be seen that the larger the value
 of $C_{s}$ is, the longer the isotropic turbulence will be developed in the top overshooting. With respect to the bottom overshooting region, the
 ratio $\overline{u_{r}^{'}u_{r}^{'}}/\overline{u_{h}^{'}u_{h}^{'}}$
 shows a much more complicated behavior for smaller values
 of $C_{s}$. When $C_{s}$ is large enough, this complication disappears and the convection becomes more and more horizontally dominated as the convection penetrates inward into the overshooting region.

   \begin{figure}[h!!!]
   \centering
      \vspace{-30pt}
   \includegraphics[width=9.6cm, angle=0]{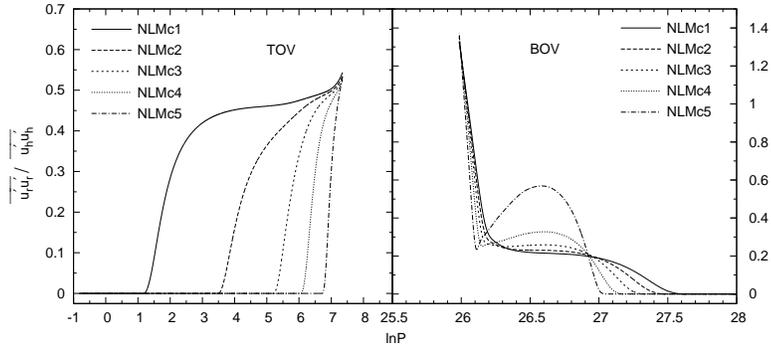}
   \begin{minipage}[]{140mm}
   \caption{Ratio of the radial turbulent kinetic energy to the horizontal one $\overline{u_{r}^{'}u_{r}^{'}}/\overline{u_{h}^{'}u_{h}^{'}}$ with different TCM's parameters for the $5M_{\odot}$ star at the AGB phase. }
   \end{minipage}
   \label{Fig11}
   \end{figure}

   \begin{figure}[h!!!]
   \flushleft
   \qquad    \quad    \qquad   \qquad
   \includegraphics[width=7.0cm, angle=0]{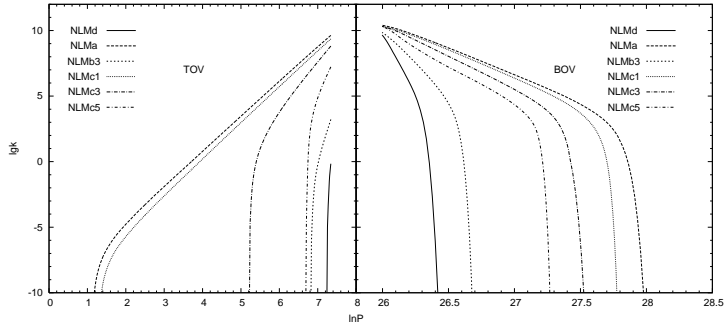}
   \begin{minipage}[]{135mm}
   \caption{The decay way of the turbulent energy $k$ in the top overshooting zone (TOV) and the bottom overshooting zone (BOV) of the $5M_{\odot}$ star at the AGB phase according to some non-local models.}
   \end{minipage}
   \label{Fig12}
   \end{figure}

   \begin{figure}[h!!!]
   \flushleft
   \qquad \quad   \qquad   \qquad
   \includegraphics[width=7.0cm, angle=0]{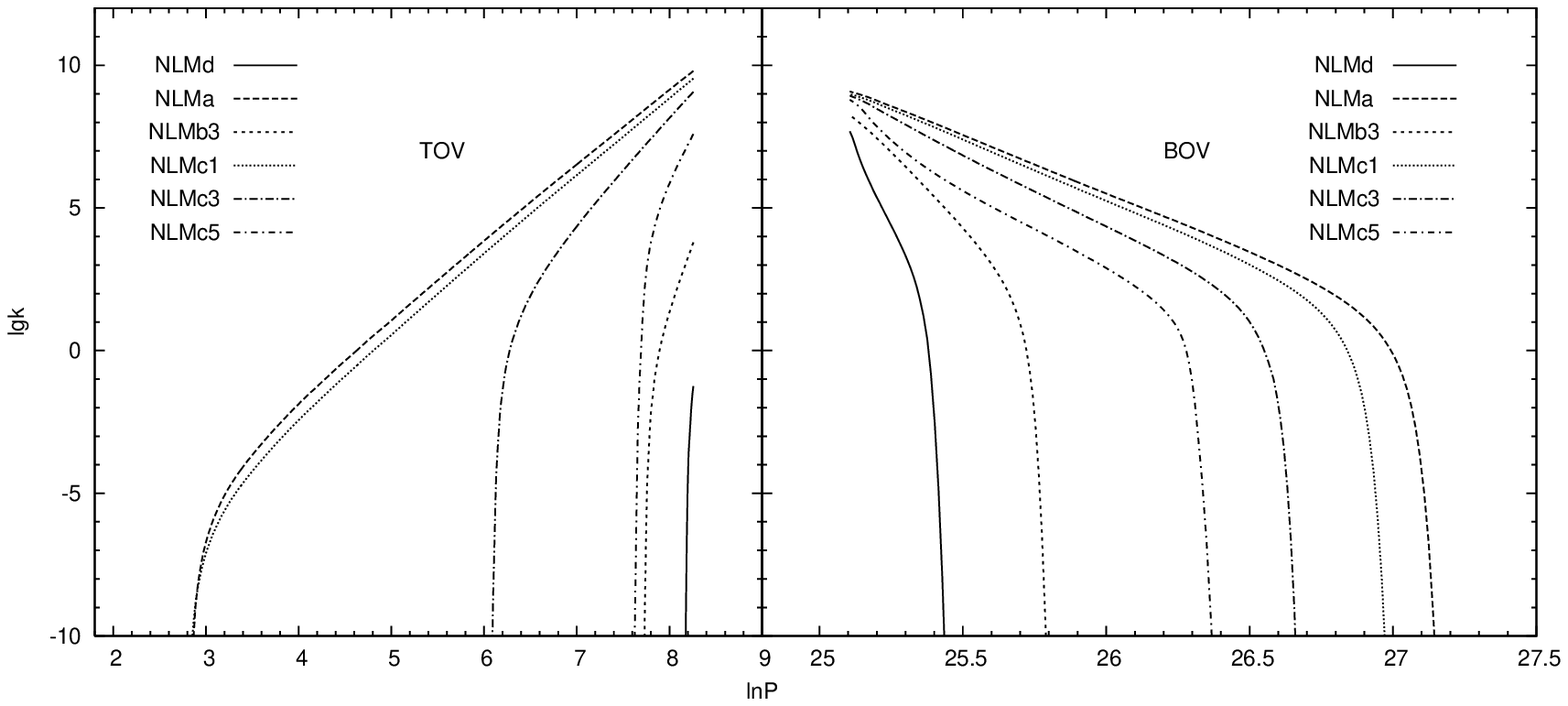}
   \begin{minipage}[]{135mm}
   \caption{ The same as Fig.12, but for the $5M_{\odot}$ star at the RGB phase.}
   \end{minipage}
   \label{Fig13}
   \end{figure}

   \begin{figure}[h!!!]
   \flushleft
   \qquad \quad   \qquad   \qquad
   \includegraphics[width=7.0cm, angle=0]{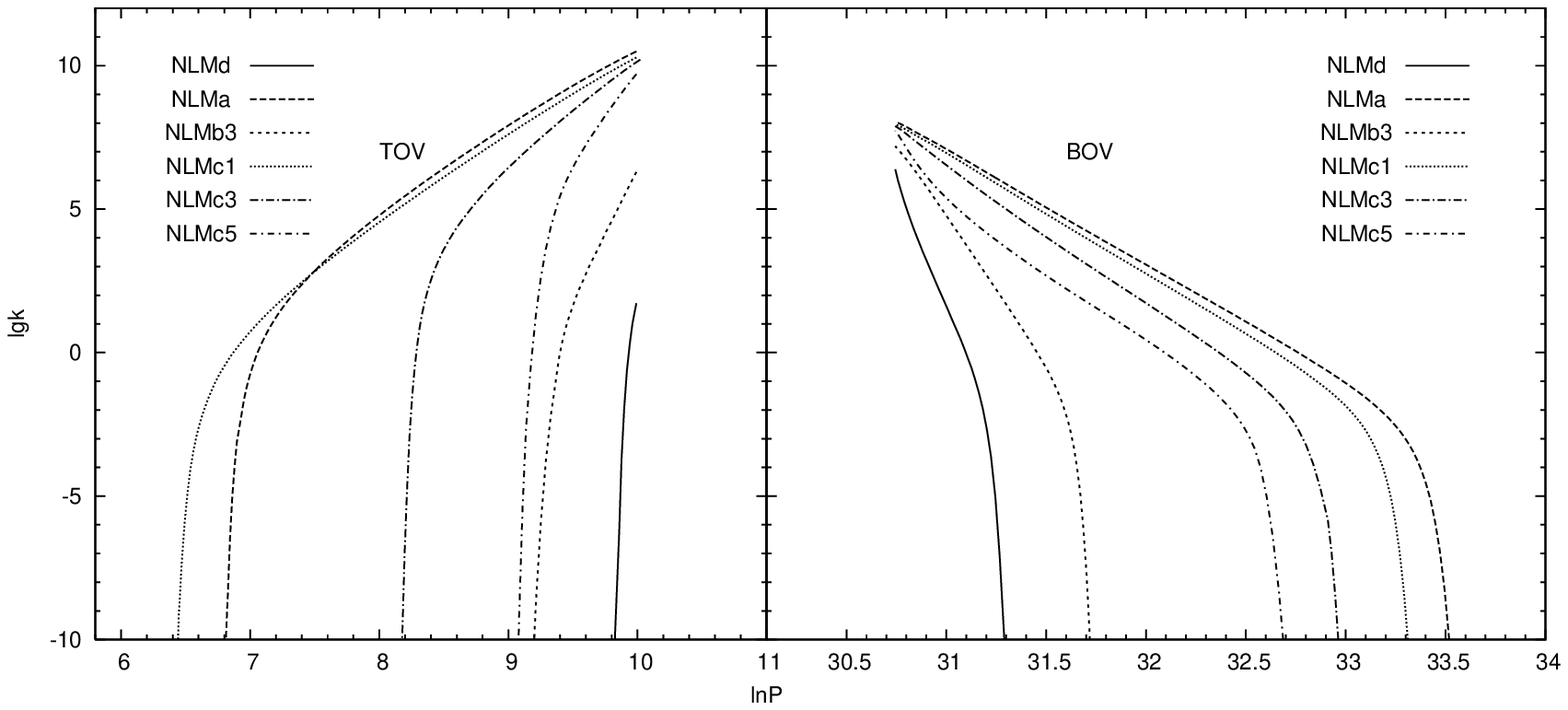}
   \begin{minipage}[]{135mm}
   \caption{ The same as Fig.12, but for the $2M_{\odot}$ star at the RGB phase. }
   \end{minipage}
   \label{Fig22}
   \end{figure}

The decaying law of the turbulent kinetic energy $k$ is a crucial
factor determining the efficiency of convective mixing in the
overshooting regions. The decay way of the turbulent kinetic energy
$k$ in the overshooting regions are given in Figs.12-14 for the
three stellar models based on different choices of the TCM's
parameters. There is a nearly linear relation between $\lg k$ and
$\ln P$ before a turning point on the curve of $\lg k$, beyond which
$k$ begins to drop promptly to zero. For example, for NLMa the
relation can be approximated as $\lg k=2.72\ln P+C$ in the top
overshooting regions, where $C=-16.8, -12.4$ and $-10.2$,
respectively, for the three stellar models up along the Hayashi
line. On the other hand, in the bottom overshooting regions we can
obtain similar relations, for instance, for NLMc3 the result is $\lg
k=-4.98\ln P+C$, where $C=161, 133.8$ and $140$ for the three
stellar models. It should be noted that the slope of the decaying
law seems only a function of the TCM's parameters, and the stellar
parameters only affect its constant.

The extension of these linearly decaying region and the slope of the
decaying law are sensitive to the TCM's parameters. It can be found
that when $C_{s}$ increases while the other parameters are fixed
(e.g. NLMc1, NLMc3, NLMc5), the turbulent kinetic energy decays
slower and slower, and the obtained linearly decaying distance will
become longer. For example, for the model of $5M_{\odot}$ star at
the AGB phase in Fig.12, the length of linearly decaying distance in
the bottom overshooting region for NLMc1, NLMc3, and NLMc5 are
respectively about $1.2H_{P}$, $1.4H_{P}$ and $1.7H_{P}$. On the
other hand, when the value of parameter $\alpha$ increases, the
slope of the decaying law becomes less and the linearly decaying
distance is significantly prolonged as seen by comparing the result
of NLMd and that of NLMa. From Figs.12-14 it can be found that the
decaying ways of the $k$ are very similar for stars of different
masses and evolutionary stages with same values of the TCM's
parameters.

\section{Concluding remarks}

Based on the TCM of Li \& Yang (2007) we have obtained the
characteristics of the turbulent convection in the RGB and AGB stars
with huge convection zones in their envelopes. Some sets of suitable
values of the TCM's parameters that have been used in the solar
convection zone are adopted to make an exploratory application in
two stars of $2$ and $5M_{\odot}$ at the RGB and AGB phases. In
practical calculations the TCM has been applied to the whole
convective envelope including the overshooting regions. When
analyzing the characteristics of the turbulent correlations, we
separate the whole convective envelope into the convective unstable
zone and the convective overshooting region. In the convective
unstable zone, we find an approximation approach to associate the
turbulent correlations explicitly, and can be used to explain the
turbulent properties. Later, we make a comparison between the
results of the TCM and those of the MLT on the velocity profile of
turbulence and in influence on stellar structure and evolution.
These obtained features of convective motion by the TCM are highly
sensitive to the TCM's parameters. The main conclusions are
summarized as follows:

\hangafter=1\setlength{\hangindent}{2.8em}1) The non-local effect of
the turbulent convection is only exhibited significantly around the
boundaries of the convective unstable zone, especially around the
upper boundary, which is in agreement with the results of Li \& Yang
(2007). The three correlations $\sqrt{k}$,
$\overline{u_{r}^{'}T^{'}}$, and $\overline{{T'}^{2}}$ are nearly
peaked at the same place in the convective unstable zone for a same
set of the TCM's parameters. It is because that all of them are
directly correlated to a function of
$(\nabla-\nabla_{ad})\overline{T}$.

\hangafter=1\setlength{\hangindent}{2.8em}2) For a fixed heat flux
(closely related by $\overline{u_{r}^{'}T^{'}}$) transferred by the
convective motion, compared to the results of the MLT, the TCM will
result in a larger turbulent velocity and thus more violent
convective motion. Furthermore, for a nearly same profile of the
convection velocity obtained by adjusting the value of the mixing
length parameter $\alpha$ for both the TCM and MLT, the
superadiabatic convection zone will be extended much more inward for
the TCM than for the case of the MLT and the stellar models based on
the TCM will be of lower effective temperature.

\hangafter=1\setlength{\hangindent}{2.8em}3) For the stellar models
located up along the Hayashi line, we find that the convective
motions become stronger and stronger. However, the e-folding lengths
of $\sqrt{k}$ in both the top and bottom overshooting regions
decrease as the stellar model is located up along the Hayashi line.
The overshooting length of $\overline{u_{r}^{'}T^{'}}$ is found to
be shorter than other turbulent quantities for a same set of the
TCM's parameters in both overshooting regions, which is consistent
with the results of Deng \& Xiong (2008) and Zhang \& Li (2009). And
the overshooting distances of $\overline{u_{r}^{'}T^{'}}$ in the
bottom overshooting region are almost the same for different stellar
models with a same set of the TCM's parameters. Therefore, the
bottom convective overshooting has a similar effect on the stellar
structure for all of the considered stellar models.

\hangafter=1\setlength{\hangindent}{2.8em}4) The diffusive parameter
$C_{s}$ plays an important role in the development of the isotropic
turbulence in the top overshooting region, and the longer the
isotropic turbulence will be obtained for larger  $C_{s}$. In the
bottom overshooting region, the convection becomes more and more
horizontally dominated for larger value of $C_{s}$, and the
convection shows a much more complicated behavior when $C_{s}$ is
small enough.

\hangafter=1\setlength{\hangindent}{2.8em}5) The decaying ways of
the turbulent kinetic energy $k$ are very similar for the different
stellar models based on a same set of TCM's parameters. We find that
there is a nearly linear relation between $\lg k$ and $\ln P$ in
most of the overshooting regions. The slope of the decaying law
seems only a function of the TCM's parameters. The values of $C_{s}$
and $\alpha$ have a similar effect on the decaying distance of the
turbulent kinetic energy, and the larger the values of them the
slope of the decaying law becomes smaller and therefore the linearly
decaying distance becomes longer.

\normalem
\begin{acknowledgements}
This work is in part supported by the National Natural Science
Foundation of China (Grant No. 10973035 and No. 10673030) and by the
Knowledge Innovation Key Program of the Chinese Academy of Sciences
under Grant No. KJCX2-YW-T24. Fruitful discussions with Q.-S. Zhang,
C.-Y. Ding and J. Su are highly appreciated.
\end{acknowledgements}

\label{lastpage}


\begin{thebibliography}{99}
\small \setlength{\itemindent}{-3mm} \setlength{\itemsep}{-0.5mm}
\setlength{\baselineskip}{4.5mm}



\bibitem[{Alexander \& Ferguson}(1994)]{Alexander+94} Alexander, D. R., \& Ferguson, J. W. 1994, \apj, 437, 879

\bibitem[{Bahcall}{et~al.}(1995)]{Bahcall+95} Bahcall, J. N., Pinsonneault, M. H., \& Wasserburg, G. J. 1995, Rev. Mod.
Phys., 67, 781

\bibitem[{Baker \& Kuhfuss}(1987)]{Baker+87} Baker, N. H., \& Kuhfuss, R. 1987, \aap, 185, 117

\bibitem[{B\"{o}hm-Vitense }(1953)]{Bohm-Vitense+53} B\"{o}hm-Vitense, E. 1953, Z. Astrophys., 32, 135

\bibitem[{B\"{o}hm-Vitense }(1958)]{Bohm-Vitense+58} B\"{o}hm-Vitense, E. 1958, Z. Astrophys., 46, 108


\bibitem[{Bressan}{et~al.}(1981)]{Aet+81} Bressan, A. G., Bertelli, G., \& Chiosi, C. 1981, \aap, 102, 25


\bibitem[{Busso}{et~al.}(2007)]{bet+07} Busso, M., Wasserburg, G. J., Nollett, K. M., \& Calandra, A. 2007, \apj, 671, 802


\bibitem[{Canuto \& Mazzitelli}(1996)]{Canuto+96} Canuto, V. M., Goldman, I., \& Mazzitelli, I. 1996, \apj, 473, 550


\bibitem[{Canuto }(1992)]{Canuto+92} Canuto, V. M. 1992, \apj, 392, 218

\bibitem[{Canuto }(1992)]{Canuto+93} Canuto, V. M. 1993, \apj, 416,
331

\bibitem[{Canuto }(1994)]{Canuto+94} Canuto, V. M. 1994, \apj, 428, 729

\bibitem[{Canuto }(1997)]{Canuto+97} Canuto, V. M. 1997, \apj, 482, 827

\bibitem[{Canuto \& Dubovikov}(1998)]{Canuto+98} Canuto, V. M., \& Dubovikov, M. S. 1998, \apj, 493,
834


\bibitem[{Canuto \& Mazzitelli}(1991)]{Canuto+91} Canuto, V. M., \& Mazzitelli, I. 1991, \apj, 370, 295


\bibitem[{Castellani}{et~al.}(1985)]{bet+85} Castellani, V., Chieffi, A., Pulone, L., \& Tornam\`{e}, A. 1985, \apj, 296, 204



\bibitem[{Castellani}{et~al.}(1971)]{cet+71} Castellani, V., Giannone, P., \& Renzini, A. 1971, Ap\&SS, 10,
340

\bibitem[{Charbonnel }(1995)]{Charbonnel+95} Charbonnel, C. 1995, \apjl, 453, L41


\bibitem[{Deng \& Xiong}(2008)]{Deng+08} Deng, L., \& Xiong, D.-R. 2008, MNRAS, 386, 1979

\bibitem[{Deng}{et~al.}(2006)]{det+06} Deng, L., Xiong, D.-R., \& Chan, K. L. 2006, \apj, 643,
426


\bibitem[{Deng \& Xiong}(2001)]{Deng+01} Deng, L. C., \& Xiong, D.-R. 2001, \chjaa, 1, 50

\bibitem[{Freytag}{et~al.}(1996)]{fet+96} Freytag, B., Ludwig, H. G., \& Steffen, M. 1996, \aap, 313, 497

\bibitem[{Gratton}{et~al.}(2000)]{get+00} Gratton, R. G., Sneden, C., Carretta, E., \& Bragaglia, A. 2000, \aap, 354, 169


\bibitem[{Gratton}{et~al.}(2004)]{get+04} Gratton, R., Sneden, C., \& Carretta, E. 2004, ARA\&A, 42, 385

\bibitem[{Herwig }(1997)]{Herwig+97} Herwig, F. 1997, \aap, 324, L81


\bibitem[{Herwig }(2005)]{Herwig+205} Herwig, F. 2005, ARA\&A, 43, 435

\bibitem[{Herwig}{et~al.}(1997)]{het+97} Herwig, F., Bl\"{o}cker, T., Sch\"{o}nberner, D., \& El Eid, M. 1997, \aap, 324, L81


\bibitem[{Hollowell \& Iben}(1988)]{Hollowell+88} Hollowell, D., \& Iben, I. Jr. 1988, \apj, 333, L25


\bibitem[{Hossain \& RodiW}(1982)]{Hollowell+82} Hossain, M. S., \& Rodi, W., 1982, Turbulent Buoyant Jets
and Plumes, ed. W. Rodi (Oxford: Pergamon Press), 121

\bibitem[{Iben}(1975)]{Iben+75} Iben, I. Jr. 1975, \apj, 196, 525

\bibitem[{Iben}(1977)]{Iben+77} Iben, I. Jr. 1977, \apj, 217, 788

\bibitem[{Iben}(1981)]{Iben+81} Iben, I. Jr. 1981, \apj, 246, 278


\bibitem[{Iglesias \& Iben}(1996)]{Iglesias+96} Iglesias, C. A., \& Rogers, F. J. 1996, \apj, 464, 943


\bibitem[{Jiang \& Iben}(1997)]{Jiang+97} Jiang, S.-Y., \& Huang, R.-Q. 1997, \aap, 317, 121


\bibitem[{Kraft}{et~al.}(1993)]{ket+93} Kraft, R. P., Sneden, C., Langer, G. E., \& Shetrone, M. D. 1993, \aj, 106, 1490

\bibitem[{Lattanzio}(1989)]{Lattanzio+89} Lattanzio, J.C. 1989, \apj, 344, L25

\bibitem[{Li \& Yang}(2001)]{Li+01} Li, Y., \& Yang, J.-Y. 2001, \chjaa, 1, 66

\bibitem[{Li \& Yang}(2007)]{Li+07} Li, Y., \& Yang, J.-Y. 2007, MNRAS, 375, 388

\bibitem[{Maeder}(1975)]{Maeder+75} Maeder, A. 1975, \aap, 40, 303


\bibitem[{Pederson}{et~al.}(1990)]{pet+90} Pederson, B. B., VandenBerg, D. A., \& Irwin, A. W. 1990, \apj, 352,
279


\bibitem[{Pilachowski}{et~al.}(1993)]{pet+93} Pilachowski, C.A., Sneden C., \& Booth, J. 1993, \apj, 407, 699

\bibitem[{Ram\'{\i}rez \& Cohen}(2002)]{Ramirez+02} Ram\'{\i}rez, S. V., \& Cohen, J. G. 2002, \aj, 123, 3277


\bibitem[{Recio-Blanco \& de Laverny}(2007)]{Recio-Blanco+07} Recio-Blanco, A., \& de Laverny, P. 2007, \aap, 461, L13

\bibitem[{Renzini}(1987)]{Renzini+87} Renzini, A. 1987, \aap, 188, 49


\bibitem[{Rogers}(1994)]{Rogers+94} Rogers, F. J., 1994, in IAU Colloquium 147, The Equation of State in Astrophysics, eds. G. Chabrier, \& E. Schatzman (Cambridge:
Cambridge Univ. Press), 16

\bibitem[{Rogers\& Iglesias}(1995)]{Rogers+95} Rogers, F. J., \& Iglesias, C. A. 1995, in ASP. Conf. Ser. 78, Astrophysical Applications of Powerful
New Databases, eds. S. J. Adelman, \& W. L. Wiese (San Francisco:
ASP), 78

\bibitem[{Rogers}{et~al.}(1996)]{ret+96} Rogers, F. J., Swenson, F. J., \& Iglesias, C. A. 1996, \apj, 456, 902

\bibitem[{Sackmann}(1980)]{Sackmann+80} Sackmann, I.J. 1980, \apj, 241, L37


\bibitem[{Salasnich}{et~al.}(1999)]{ret+99} Salasnich, B., Bressan, A., \& Chiosi, C. 1999, \aap, 342, 131

\bibitem[{Saslaw \& Schwarzschild}(1965)]{Saslaw+65} Saslaw, W. C., \& Schwarzschild, M. 1965, \apj, 142, 1468

\bibitem[{Spruit}{et~al.}(1990)]{set+90} Spruit, H. C., Nordlund, A., \& Title, A. M. 1990, ARA\&A, 28,
263


\bibitem[{Sreenivasan \& Wilson}(1978)]{Sreenivasan+78} Sreenivasan, S. R., \& Wilson, W. J. F. 1978, Ap\&SS, 53, 193


\bibitem[{Straniero}{et~al.}(1997)]{set+97} Straniero, O., Chieffi, A., Limongi, M., et al. 1997, \apj, 478, 332


\bibitem[{Ulrich}(1970)]{Ulrich+70} Ulrich, R. K. 1970, Ap\&SS, 7, 183

\bibitem[{Ventura \& D'Antona}(2005)]{Ventura+05} Ventura, P., \& D'Antona, F. 2005, \aap, 431, 279


\bibitem[{Xiong}(1985)]{Xiong+85} Xiong, D.-R. 1985, \aap, 150, 133

\bibitem[{Xiong}(1990)]{Xiong+90} Xiong, D.-R. 1990, \aap, 232, 31

\bibitem[{Xiong}(1979)]{Xiong+79} Xiong, D.-R. 1979, Acta Astron. Sin., 20, 238

\bibitem[{Xiong}{et~al.}(1997)]{xet+97} Xiong, D.-R., Cheng, Q.-L., \& Deng, L. 1997, \apjs, 108, 529

\bibitem[{Xiong}(1979)]{Xiong+81} Xiong, D.-R. 1981, Sci. Sinica., 24, 1406

\bibitem[{Yang \& Li}(2007)]{Yang+07} Yang, J.-Y., \& Li, Y. 2007, MNRAS, 375, 403

\bibitem[{Zhang \& Li}(2009)]{Yang+07} Zhang, Q.-S., \& Li, Y. 2009, RAA, 9, 585

\end{thebibliography}
\end{document}